\documentclass[notitlepage]{revtex4-1}
\usepackage{setspace}
\usepackage[utf8]{inputenc}
\usepackage[dvipsnames]{xcolor}
\usepackage[colorlinks=true,citecolor=RedViolet,urlcolor=BlueGreen,linkcolor=RedViolet]{hyperref}
\usepackage[normalem]{ulem}
\usepackage{url}
\usepackage{graphicx,wrapfig,float,slashed,cancel}
\usepackage{amsmath,amssymb,epsfig,xcolor,stmaryrd}
\usepackage{bm}
\usepackage{enumitem}
\usepackage{hhline,multirow,tabularx} 
\usepackage{graphics}
\usepackage{latexsym}
\usepackage{rotating}
\usepackage{subfigure}
\usepackage{color}
\usepackage{blindtext}
\usepackage{lipsum}
\usepackage{multirow}
\usepackage{physics}
\usepackage{orcidlink}

\allowdisplaybreaks
\begin{document}
	
	
	\title{\textcolor{BlueViolet}{Semileptonic decay of the triply heavy $\Omega_{ccb}$ to the observed $\Xi^{++}_{cc}$ state}}

	\author{Z.~Rajabi Najjar$^{a}$\orcidlink{0009-0002-2690-334X}}
	\email{rajabinajar8361@ut.ac.ir }
	
	\author{K.~ Azizi$^{a,b}$\orcidlink{0000-0003-3741-2167}}
	\email{kazem.azizi@ut.ac.ir}
	\thanks{Corresponding author}
	\author{H.~R.~Moshfegh$^{a,c}$\orcidlink{0000-0002-9657-7116}}
	\email{hmoshfegh@ut.ac.ir }

	\affiliation{
		$^{a}$Department of Physics, University of Tehran, North Karegar Avenue, Tehran 14395-547, Iran\\
		$^{b}$Department of Physics, Do\v{g}u\c{s} University, Dudullu-\"{U}mraniye, 34775 Istanbul, T\"urkiye\\
		$^{c}$Centro Brasileiro de Pesquisas F´ısicas, Rua Dr. Xavier Sigaud,150, URCA, Rio de Janeiro CEP 22290-180, RJ, Brazil
	}

	\date{\today}
	
	\preprint{}
	
	\begin{abstract}
	We investigate  the weak semileptonic decay of the  $ \Omega^{+}_{ccb} \rightarrow \Xi^{++}_{cc}  ~{\ell}\bar\nu_{\ell}$, where a triply heavy baryon with spin 1/2 decays into the observed doubly heavy baryon with spin 1/2, using QCD sum rule method in  all lepton channels. We compute the six relevant vector and axial vector form factors entering the low energy matrix elements in full theory. The invariant form factors are building blocks, using the fit faction of which in terms of $ q^2 $ in whole physical region, we calculate the exclusive widths  in three lepton channels.  Our predictions may  help the present and future experiments in the course of their search for doubly and triply heavy baryons.

	\end{abstract}
	
	
	\maketitle
	
	\renewcommand{\thefootnote}{\#\arabic{footnote}}
	\setcounter{footnote}{0}
	
	\section {Introduction}\label{sec:one}
Research on heavy baryons has gained significant momentum in the last decade due to  notable experimental discoveries in the spectroscopy of heavy hadrons. These advancements have driven considerable progress in the field. Among the various types of heavy baryons, those containing one heavy quark ($b$, $c$) have been extensively investigated and observed in numerous  experimental facilities. In 2017, the LHCb collaboration reported the discovery of the doubly charmed baryon, $\Xi_{cc}^{++}$, well described by the quark model in its lowest state of doubly heavy baryons, in the final state $\Lambda_c^{+}K^{-}\pi^+\pi^+$\cite{LHCb:2017iph}. Subsequently, they measured the lifetime of  $\Xi_{cc}^{++}$~\cite{LHCb:2018zpl} and identified  another decay mode,  $\Xi_{cc}^{++}\to\Xi_{c}^{+}\pi^{+}$
	\cite{LHCb:2018pcs}. In 2022, LHCb  announced the observation of a new channel $\Xi_{cc}^{++}\to\Xi_{c}^{\prime+}\pi^{+}$ from an experimental perspective~\cite{LHCb:2022rpd}. Triply heavy baryons, consisting of three heavy quarks (b,c), represent the last missing category of standard hadrons. Following the discovery of the $\Xi^{++}_{cc}$, the detection of triply heavy baryons now seems increasingly likely. While theoretical studies of triply heavy baryons have largely focused on their mass spectra \cite{Roberts:2007ni,Patel:2008mv,Vijande:2015faa,Shah:2017jkr,Shah:2018div,Shah:2018bnr,Liu:2019vtx,Migura:2006ep,Martynenko:2007je,Yang:2019lsg,Faustov:2021qqf,Meinel:2010pw,Meinel:2012qz,Briceno:2012wt,Padmanath:2013zfa,Brown:2014ena,Mathur:2018epb,Can:2015exa,Zhang:2009re,Wang:2011ae,Aliev:2012tt,Aliev:2014lxa,Wang:2020avt,Hasenfratz:1980ka,Bernotas:2008bu,Wei:2015gsa,Wei:2016jyk,Oudichhya:2021yln,Oudichhya:2021kop,Oudichhya:2023pkg,Radin:2014yna,Gutierrez-Guerrero:2019uwa,Qin:2019hgk,Yin:2019bxe,Zhao:2023qww,Silvestre-Brac:1996myf,Jia:2006gw,Brambilla:2009cd,Llanes-Estrada:2011gwu,Thakkar:2016sog,Serafin:2018aih,Shah:2023zph,Zhou:2024fjj,Najjar:2024deh,Oudichhya:2023pkg,Xie:2024lfo},  relatively less attention has been given to their decay properties  \cite{Flynn:2011gf, Geng:2017mxn,Wang:2018utj, Huang:2021jxt,Zhao:2018mrg,Wang:2022ias,Lu:2024bqw}. Decay processes discussed in the literature predominantly involve transitions from spin 3/2 to final states with spin 1/2  \cite{ Geng:2017mxn,Wang:2018utj,Huang:2021jxt,Zhao:2018mrg,Wang:2022ias,Lu:2024bqw}, or from spin 1/2 to 3/2 \cite{Flynn:2011gf}.  
	
	The present  study focuses on the investigation of the semileptonic decay of the  triply heavy baryon $ \Omega^{+}_{ccb}  $ to the doubly heavy baryon $  \Xi^{++}_{cc}  $, where both the initial and final particles have spin 1/2.  To find the relevant form factors and estimate the decay width for the process $ \Omega^{+}_{ccb}\rightarrow \Xi^{++}_{cc}  ~{\ell}\bar\nu_{\ell}$ in all lepton channels, we employ the QCD sum rule method, introduced  by Shifman, Vainshtein, and Zakharov \cite{Shifman:1978bx, Shifman:1978by}, a well-established nonperturbative approach recognized for its effectiveness in predicting hadronic properties \cite{Aliev:2009jt,Aliev:2010uy,Aliev:2012ru,Agaev:2016mjb,Azizi:2016dhy}. This methodology evaluates  the  decay  from two distinct perspectives: the first is the physical side, where results are  expressed in terms of the hadronic parameters like mass and  residue of  both the initial and final baryons as well as the transition form factors; the second is the QCD side, where results are derived in terms of the  fundamental degrees of freedom like quark masses, quark-gluon coupling constant and   condensates associated with the quark-quark, gluon-gluon, and quark-gluon interactions through the QCD vacuum. The form factors are subsequently computed by correlating the outcomes of these two approaches, utilizing principles of quark-hadron duality assumption,  double Borel transformation as well as continuum subtraction. Ultimately,  the obtained form factors  are used  to determine the relevant semileptonic decay widths at different lepton channels.
	
	 The structure of the rest of this article is  as follows: Section \ref{Sec:two}  outlines the formalism of QCD sum rules used to calculate an appropriate correlation function in  both the physical and QCD sectors. We derive the desired sum rules for the form factors based on the standard prescription of the method. Section  \ref{Sec:three} is dedicated to the numerical analysis of the form factors. In section \ref{Sec:four}, we present numerical results for  the semileptonic decay widths across all lepton channels.  Section \ref{Sec:five} includes our concluding remarks  and some details of the calculations are presented in the Appendices.

	\section {Formalism }\label{Sec:two} 
\begin{figure}[h!] 
		\includegraphics[totalheight=3.5cm,width=6cm]{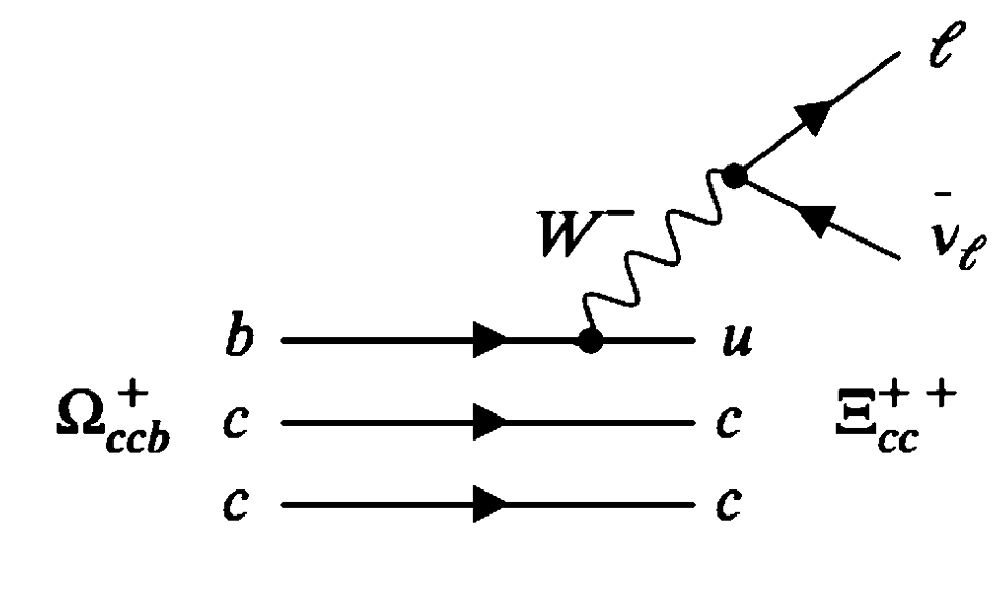}
		\caption{The $ \Omega_{ccb}^{+}\rightarrow \Xi^{++}_{cc}  ~{\ell}\bar\nu_{\ell}$ semileptonic decay.}\label{Fig:current}
	\end{figure}

	In the semileptonic decay $ \Omega_{ccb}^{+}\rightarrow \Xi^{++}_{cc}  ~{\ell}\bar\nu_{\ell}$ decay, the  charm quarks  act as the spectators and the transition is conducted by the weak current ${\cal
		J}_{\mu}^{tr}=\bar u \gamma_\mu(1-\gamma_5) b$ through the process $b\to u {\ell}\bar\nu_{\ell}$ at quark level (see Fig.~\ref{Fig:current}). The  effective Hamiltonian  describing this process is given by:
	\begin{eqnarray}\label{Heff}
	{\cal H}_{eff} =
	\frac{G_F}{\sqrt2} V_{ub} ~\bar u \gamma_\mu(1-\gamma_5) b \bar{\ell}\gamma^\mu(1-\gamma_5) \nu_{\ell},
\end{eqnarray}
where $G_F$ represents the Fermi coupling constant and $V_{ub}$ is the Cabibbo-Kobayashi-Maskawa (CKM) matrix element. By sandwiching the effective Hamiltonian between the initial triply  and the final doubly heavy baryon states, the corresponding amplitude is obtained: 
\begin{eqnarray}\label{amp}
	M=\langle \Xi^{++}_{cc}\vert{\cal H}_{eff}\vert \Omega_{ccb}^{+}\rangle 
=\frac{G_F}{\sqrt2} V_{ub}\bar{\ell}~\gamma^\mu(1-\gamma_5) \nu_{\ell} \langle \Xi^{++}_{cc}\vert \bar u \gamma_\mu(1-\gamma_5)b\vert \Omega_{ccb}^{+}\rangle.
\end{eqnarray}
As is seen, the transition current  consists of  two components: the vector ($V^\mu=\bar u \gamma_\mu b$) and the axial vector ($A^\mu=\bar u \gamma_\mu\gamma_5b$), each of which contains information about three transition form factors. 
Therefore, the following  low energy matrix elements  are defined in terms of six form factors by  taking into account the  Lorentz invariance and parity considerations:
\begin{eqnarray}\label{Cur.with FormFac.}
	&&\langle \Xi^{++}_{cc}(p',s')|V^{\mu}|\Omega_{ccb}^{+} (p,s)\rangle = \bar
	u_{\Xi^{++}_{cc}}(p',s') \Big[F_1(q^2)\gamma^{\mu}+F_2(q^2)\frac{p^{\mu}}{m_{\Omega_{ccb}^{+}}}
	+F_3(q^2)\frac{p'^{\mu}}{m_{\Xi^{++}_{cc}}}\Big] u_{\Omega_{ccb}^{+}}(p,s), \notag \\
	&&\langle \Xi^{++}_{cc}(p',s')|A^{\mu}|\Omega_{ccb}^{+} (p,s)\rangle = \bar u_{\Xi^{++}_{cc}}(p',s') \Big[G_1(q^2)\gamma^{\mu}+G_2(q^2)\frac{p^{\mu}}{m_{\Omega_{ccb}^{+}}}+G_3(q^2)\frac{p'^{\mu}}{m_{\Xi^{++}_{cc}}}\Big]
	\gamma_5 u_{\Omega_{ccb}^{+}}(p,s), 
\end{eqnarray}
where $F_i$ and $G_i$ ( $i$=1,2,3) represent the form factors associated with the vector and axial components within the full  framework, respectively. The four-momenta of the initial and final baryons are denoted as $p$ and $p'$, respectively, while $q = p - p'$ represents the four-momentum transferred to the lepton pair. The symbols $u_{\Omega_{ccb}^{+}}(p,s)$  and $u_{\Xi^{++}_{cc}}(p',s')$  correspond to the Dirac spinors of the initial and final states. As mentioned, the form factors are fundamental parameters that can be extracted utilizing the QCD sum rule method. The first step in this process involves selecting an appropriate  correlation function. To derive the form factors, we express the three-point correlation function as follows:
\begin{eqnarray}\label{CorFunc}
	\Pi_{\mu}(p,p^{\prime},q)&=&i^2\int d^{4}x e^{-ip\cdot x}\int d^{4}y e^{ip'\cdot y}  \langle 0|{\cal T}\{{\cal J}^{\Xi^{++}_{cc}}(y){\cal
		J}_{\mu}^{tr,V(A)}(0) \bar {\cal J}^{\Omega_{ccb}^{+}}(x)\}|0\rangle,
\end{eqnarray}
where the symbol $\mathcal{T}$ denotes the time-ordered product, while ${\cal
	J}_{\mu}^{tr,V(A)}(0)$ represents the transition current of semileptonic decay.  $ {\cal J}^{\Omega_{ccb}^{+}}(x)$ and ${\cal J}^{\Xi_{cc}^{++}}(y) $ signify the interpolating currents of the initial and final particles, respectively. In this case, the initial particle is a triply heavy baryon with spin 1/2, and the final particle is a symmetric doubly heavy spin 1/2 baryon. The general expressions of the interpolating currents for these specified particles are given as follows:
\begin{equation}\label{cur1}
	{\cal J}^{\Omega_{QQQ'}}(x)=2\varepsilon^{abc}\Big\{\Big(Q^{aT}(x)CQ^{'b}(x)\Big)\gamma_{5}Q^c(x)+
	\beta\Big(Q^{aT}(x)C\gamma_{5}Q^{'b}(x)\Big)Q^c (x)\Big\}, 
\end{equation}
and
\begin{align}\label{eq:CorrF2}
	{\cal J}^{\Xi_{QQ}}(x)=\frac{2}{\sqrt{2}} \epsilon^{abc} \Big\{\Big(Q^{a^T}(x) C q^b(x)\Big)\gamma_5 Q^c(x)  +\beta\Big(Q^{a^T}(x) C \gamma_5 q^b(x)\Big) Q^c(x)\Big\}.
\end{align}	
Here the superscripts $a$, $b$, and $c$ denote color indices, $C$ represents the charge conjugation operator, and $\beta$ is an arbitrary auxiliary parameter. The variable $q$ correspond to light quark, $u$, while $Q^{(')}$ denotes the heavy quarks $c(b)$. Within the framework of QCD sum rules, the correlation function must be computed from both the hadronic and the QCD  sides. Then the form factors can be achieved by relating the results derived from these two regimes, applying dispersion integrals and quark-hadron duality assumption and utilizing the Borel transformation technique and
continuum subtraction to suppress  contributions of higher exited states.
	\subsection {Hadronic side }\label{subSec:one}
		In this subsection, the correlation function is evaluated within the timelike region of the light cone. To achieve this, it is imperative to incorporate two relevant complete sets of states that possess the same quantum numbers as the initial,  $\Omega_{ccb}^{+}$, and final, $\Xi^{++}_{cc}$, states at the appropriate stage. Subsequently, by performing the integrals over four dimensional coordinates $ x $ and $ y $, the form of the correlation function in the hadronic sector is derived as follows:
		
	\begin{eqnarray} \label{PhysSide}
		\Pi_{\mu}^{Had.}(p,p',q)=\frac{\langle 0 \mid {\cal J}^{\Xi^{++}_{cc}} (0)\mid \Xi^{++}_{cc}(p') \rangle \langle \Xi^{++}_{cc} (p')\mid
			{\cal J}_{\mu}^{tr,V(A)}(0)\mid \Omega_{ccb}^{+}(p) \rangle \langle \Omega_{ccb}^{+}(p)
			\mid\bar {\cal J}^{\Omega_{ccb}^{+}}(0)\mid
			0\rangle}{(p'^2-m_{\Xi^{++}_{cc}}^2)(p^2-m_{\Omega_{ccb}^{+}}^2)}+\cdots,
	\end{eqnarray}
	where the first term is the ground state contribution  and
dots refer to the  contributions from higher resonances and continuum modes. The hadronic matrix elements referenced in the  correlation function are defined by the subsequent equations:
\begin{eqnarray}\label{MatrixElements}
	&&\langle 0|{\cal J}^{\Xi^{++}_{cc}}(0)|\Xi^{++}_{cc}(p')\rangle =
	\lambda_{\Xi^{++}_{cc}} u_{\Xi^{++}_{cc}}(p',s'), \notag \\
	&&\langle\Omega_{ccb}^{+}(p)|\bar {\cal J}^{\Omega_{ccb}^{+}}(0)| 0 \rangle =
	\lambda^{\dag}_{\Omega_{ccb}^{+}}\bar u_{\Omega_{ccb}^{+}}(p,s),
\end{eqnarray}
the symbols $\lambda_{\Omega_{ccb}^{+}}$ and $\lambda_{\Xi^{++}_{cc}}$ correspond to the residue of  the  initial  and the final states , respectively. $u_{\Omega_{ccb}^{+}}(p,s)$ and $u_{\Xi^{++}_{cc}}(p',s')$   satisfy  the following identities: 
\begin{eqnarray}\label{Spinors}
	\sum_{s'} u_{\Xi^{++}_{cc}} (p',s')~\bar{u}_{\Xi^{++}_{cc}}
	(p',s')&=&\slashed{p}~'+m_{\Xi^{++}_{cc}},\notag \\
	\sum_{s} u_{\Omega_{ccb}^{+}}(p,s)~\bar{u}_{\Omega_{ccb}^{+}}(p,s)&=&\slashed
	p+m_{\Omega_{ccb}^{+}}.
\end{eqnarray}
By substituting the lately straightforward mathematical relations into Eq.~(\ref{PhysSide}), the general form of the correlation function in the hadronic sector can be expressed as follows:
\begin{eqnarray} \label{PhysSidetotal}
	\Pi_{\mu}^{Had.}(p,p',q)&=&\frac{\lambda_{\Xi^{++}_{cc}}\lambda^{\dag}_{\Omega_{ccb}^{+}}(\slashed{p}~'+m_{\Xi^{++}_{cc}})(\slashed{p}~+m_{\Omega_{ccb}^{+}})}{(p'^2-m_{\Xi^{++}_{cc}}^2)(p^2-m_{\Omega_{ccb}^{+}}^2)}\Bigg( \Big[F_1(q^2)\gamma^{\mu}+F_2(q^2)\frac{p^{\mu}}{m_{\Omega_{ccb}^{+}}}
		+F_3(q^2)\frac{p'^{\mu}}{m_{\Xi^{++}_{cc}}}\Big]\notag\\ &-&\Big[G_1(q^2)\gamma^{\mu}\gamma_5+G_2(q^2)\frac{p^{\mu}\gamma_5}{m_{\Omega_{ccb}^{+}}}+G_3(q^2)\frac{p'^{\mu}\gamma_5}{m_{\Xi^{++}_{cc}}}\Big]
		\Bigg)+\cdots.
\end{eqnarray}
To suppress the contributions from excited and continuum states, we employ the double Borel transformation:
\begin{eqnarray}\label{BorelQCD2}
	\mathbf{\widehat{B}}\frac{1}{(p^{2}-s)^m} \frac{1}{(p'^{2}-s')^n}\longrightarrow (-1)^{m+n}\frac{1}{\Gamma[m]\Gamma[n]} \frac{1}{(M^2)^{m-1}}\frac{1}{(M'^2)^{n-1}}e^{-s/M^2} e^{-s'/M'^2},
\end{eqnarray}
in this transformation,  $M^2$ and $M'^2$ represent Borel parameters, which are determined by establishing  working windows in the numerical analysis section. In the Borel scheme, we have

\begin{eqnarray} \label{PhysSidetotalBorel}
	&\mathbf{\widehat{B}}&~\Pi_{\mu}^{Had.}(p,p',q)=\lambda_{\Xi^{++}_{cc}}\lambda^{\dag}_{\Omega_{ccb}^{+}}(\slashed{p}~'+m_{\Xi^{++}_{cc}})(\slashed{p}~+m_{\Omega_{ccb}^{+}})e^{-\frac{m_{\Omega_{ccb}^{+}}^2}{M^2}}
	e^{-\frac{m_{\Xi^{++}_{cc}}^2}{M'^{2}}}\\ \notag
	&\times&\Bigg( \Big[F_1(q^2)\gamma^{\mu}+F_2(q^2)\frac{p^{\mu}}{m_{\Omega_{ccb}^{+}}}
	+F_3(q^2)\frac{p'^{\mu}}{m_{\Xi^{++}_{cc}}}\Big] -\Big[G_1(q^2)\gamma^{\mu}\gamma_5+G_2(q^2)\frac{p^{\mu}\gamma_5}{m_{\Omega_{ccb}^{+}}}+G_3(q^2)\frac{p'^{\mu}\gamma_5}{m_{\Xi^{++}_{cc}}}\Big]
	\Bigg)+\cdots.~
\end{eqnarray}
\subsection {QCD side }\label{subSec:two}
As previously indicated, within the QCD sum rule formalism, it is crucial to compute the correlation function in the QCD sector in the deep Euclidean spacelike region through the application of operator product expansion (OPE). To accomplish this objective, one must first substitute the interpolating current of the initial and final particle given in Eqs.~(\ref{cur1}) and (\ref{eq:CorrF2})  within Eq.~(\ref{CorFunc}). Following this substitution, and applying Wick's theorem to compute all possible contractions of the corresponding quark fields,  a clear expression emerges that includes the  heavy and light quark propagators for the correlation function:
\begin{eqnarray}\label{eqQCD1}
	\Pi_\mu(p,p')&=&2\sqrt2 \epsilon^{abc}\epsilon^{a^{'}b^{'}c^{'}}\int d^{4}x e^{-ip\cdot x}\int d^{4}y e^{ip'\cdot y} \Big\{\gamma_5~S^{cc'}_c(y-x)~\gamma_{5}~Tr\Big[S^{'aa'}_c(y-x)~S^{bi}_u(y)~\gamma_\mu~(1-\gamma_5)~S^{ib'}_b(-x)\Big]\notag\\
		&-&\gamma_5~S^{ca'}_c(y-x)~S^{'ib'}_b(-x)~(1-\gamma_5)\gamma_\mu~S^{'bi}_u(y)~S^{ac'}_c(y-x)~\gamma_5\notag\\
		&+&\beta~\bigg(\gamma_5~S^{cc'}_c(y-x)~Tr\Big[S^{'aa'}_c(y-x)~S^{bi}_u(y)~\gamma_\mu(1-\gamma_5)~S^{ib'}_b(-x)~\gamma_5\Big]\notag\\
		& -&\gamma_5~S^{ca'}_c(y-x)~
\gamma_5~S^{'ib'}_b(-x)~(1-\gamma_5)\gamma_\mu~S^{'bi}_u(y)~S^{ac'}_c(y-x)	\notag\\
&+&S^{cc'}_c(y-x)~\gamma_5~Tr\Big[S^{'aa'}_c(y-x)~\gamma_5~S^{bi}_u(y)~\gamma_\mu(1-\gamma_5)~S^{ib'}_b(-x)\notag\\
&-&S^{ca'}	_c(y-x)~S^{'ib'}_b(-x)~(1-\gamma_5)\gamma_\mu	~S^{'bi}_u(y)~\gamma_5~S^{ac'}_c(y-x)~\gamma_5\bigg)\notag\\
		&+&\beta^2~\bigg(S^{cc'}_c~Tr\Big[S^{'aa'}_c(y-x)~\gamma_5~S^{bi}_u(y)~\gamma_\mu(1-\gamma_5)
~S^{ib'}_b(-x)~\gamma_5\Big]\notag\\
&-&S^{ca'}_c(y-x)~\gamma_5~S^{'ib'}_b(-x)~(1-\gamma_5)\gamma_\mu
~S^{'bi}_u(y)~\gamma_5~S^{ac'}_c(y-x)\bigg)		\Big\}.
\end{eqnarray}	
Here, $S_{c(b)}$ denotes the heavy quark propagator, $S_u$ represents the light quark propagator, and  $S'=CS^TC$. To conduct calculations in coordinate space, one should utilize the following explicit formulas for the  heavy and light propagators, respectively. We take:
\begin{eqnarray}\label{LightProp}
	S_{q}^{ab}(x)&=&i\delta _{ab}\frac{\slashed x}{2\pi ^{2}x^{4}}-\delta _{ab}%
	\frac{m_{q}}{4\pi ^{2}x^{2}}-\delta _{ab}\frac{\langle\overline{q}q\rangle}{12} +i\delta _{ab}\frac{\slashed xm_{q}\langle \overline{q}q\rangle }{48}%
	-\delta _{ab}\frac{x^{2}}{192}\langle \overline{q}g_{}\sigma
	Gq\rangle+
	i\delta _{ab}\frac{x^{2}\slashed xm_{q}}{1152}\langle \overline{q}g_{}\sigma Gq\rangle \notag\\
	&-&i\frac{g_{}G_{ab}^{\alpha \beta }}{32\pi ^{2}x^{2}}\left[ \slashed x{\sigma _{\alpha \beta }+\sigma _{\alpha \beta }}\slashed x\right]-i\delta _{ab}\frac{x^{2}\slashed xg_{}^{2}\langle
		\overline{q}q\rangle ^{2}}{7776} -\delta _{ab}\frac{x^{4}\langle \overline{q}q\rangle \langle
		g_{}^{2}G^{2}\rangle }{27648}+\ldots,
\end{eqnarray}
and
\begin{eqnarray}
	&&S_{Q}^{ab}(x)=i\int \frac{d^{4}k}{(2\pi )^{4}}e^{-ikx}\Bigg \{\frac{\delta
		_{ab}\left( {\slashed k}+m_{Q}\right) }{k^{2}-m_{Q}^{2}}-\frac{%
		g_{s}G_{ab}^{\alpha \beta }}{4}\frac{\sigma _{\alpha \beta }\left( {\slashed %
			k}+m_{Q}\right) +\left( {\slashed k}+m_{Q}\right) \sigma _{\alpha \beta }}{%
		(k^{2}-m_{Q}^{2})^{2}}  \notag  \label{eq:A2} \\
	&&+\frac{g_{s}^{2}G^{2}}{12}\delta _{ab}m_{Q}\frac{k^{2}+m_{Q}{\slashed k}}{%
		(k^{2}-m_{Q}^{2})^{4}}+\frac{g_{s}^{3}G^{3}}{48}\delta _{ab}\frac{\left( {%
			\slashed k}+m_{Q}\right) }{(k^{2}-m_{Q}^{2})^{6}}\left[ {\slashed k}\left(
	k^{2}-3m_{Q}^{2}\right) +2m_{Q}\left( 2k^{2}-m_{Q}^{2}\right) \right] \left(
	{\slashed k}+m_{Q}\right) +\cdots \Bigg \},  
	\end{eqnarray}  
where  $m_{Q(q)}$ denote the mass of the  heavy (light) quark  and $k$ represents the four-momentum. $G_{ab}^{\mu \nu }=G_{A}^{\mu\nu
}t_{ab}^{A}$, $t^A=\lambda^A/2$ ,  $G^{2}=G_{A}^{\mu\nu} G_{\mu \nu }^{A}$ ,  $G_{\mu\nu}$ denotes the gluon field strength tensor, $\lambda
^{A}$ illustrate the Gell-Mann matrices which $A=1,\,2\,\ldots 8$; \(\langle \overline{q}q\rangle\), \(\langle G^2\rangle\) and \(\langle \overline{q}g_{}\sigma Gq\rangle\) represent quark-quark, gluon-gluon and quark-gluon condensates, respectively. At this stage, following the applying the quark propagators in coordinate space to the correlation function, we proceed to compute the integrals within the QCD side utilizing a series of mathematical techniques. These techniques include Fourier transforms, Feynman parametrization and various identities, as detailed. An illustration of this process is provided in:
\begin{eqnarray}\label{exampleterm}
	\int d^4k_1\int d^4k_2\int d^4k_3 \int d^4x e^{i(k_1+k_2+k_3-p).x}\int d^4y e^{i(-k_1-k_2+p').y} \frac{f(k_1,k_2,k_3,y)} {(k_1^2-m^2_c)^{n_1}(k_2^2-m^2_{c})^{n_2}(k_3^2-m^2_b)^{n_3}y^{2n_4}}.
\end{eqnarray}
In the initial step, we employ the following identity  \cite{Azizi:2017ubq}:
\begin{eqnarray}\label{intyx}
	\frac{1}{y^{2n}}&=&\int\frac{d^Dt}{(2\pi)^D}e^{-it\cdot y}~i~(-1)^{n+1}~2^{D-2n}~\pi^{D/2}  \frac{\Gamma(D/2-n)}{\Gamma(n)}\Big(-\frac{1}{t^2}\Big)^{D/2-n},
\end{eqnarray}
inserting of $y_{\mu}\rightarrow
-i\frac{\partial}{\partial p'_{\mu}}$ into the aforementioned relation yields:
\begin{eqnarray}\label{exampleterm3}
&\int d^Dt&	\int d^4k_1\int d^4k_2\int d^4k_3 \int d^4x e^{i(k_1+k_2+k_3-p).x}\int d^4y e^{i(-k_1-k_2+p'-t).y}\notag\\ &\times&\frac{f(k_1,k_2,k_3)} {(k_1^2-m^2_c)^{n_1}(k_2^2-m^2_{c})^{n_2}(k_3^2-m^2_b)^{n_3}}\Big(-\frac{1}{t^2}\Big)^{D/2-n_4}.
\end{eqnarray}
In the subsequent step, we apply Fourier integrals with respect to the four variables $  x$ and $  y$:
\begin{eqnarray}\label{fourier}
	\int d^4x e^{i(k_1+k_2+k_3-p).x}\int d^4ye^{i(-k_1-k_2+p'-t).y}= (2\pi)^4\delta^4(k_1+k_2+k_3-p) (2\pi)^4\delta^4(-k_1-k_2+p'-t).
\end{eqnarray}
Making use of the two Dirac delta functions derived  in  Eq.~(\ref{fourier}),  the four-integrals over $k_2$ and $k_3$ can be easily  performed.  In the third step,  the remaining four-integrals over $k_1$
and $t$ are evaluated using the Feynman parametrization:
\begin{eqnarray}
	\label{feynman }
	\frac{1}{A_1^{n_1}A_2^{n_2}A_3^{n_3}A_4^{n_4}}&=&\frac{\Gamma(n_1+n_2+n_3+n_4)}{\Gamma({n_1})\Gamma({n_2})\Gamma({n_3})\Gamma({n_4})}\int_0^1\int_0^{1-r}\int_0^{1-r-z}dvdzdr\notag\\
	&\times&\frac{r^{n_1-1} \ z^{n_2-1}v^{n_3-1}(1-r-z-v)^{n_4-1}}{[rA_1 \ +zA_2 \ +vA_3 \ +(1-r-z-v)A_4]^{n_1+n_2+n_3+n_4}}.
\end{eqnarray} 
In the fourth step, we use the following identity to address the remaining integrals \cite{Azizi:2017ubq}:
\begin{eqnarray}\label{Int}
	\int d^Dt\frac{(t^2)^{m}}{(t^2+\Delta)^{n}}=\frac{i \pi^2
		(-1)^{m-n}\Gamma(m+2)\Gamma(n-m-2)}{\Gamma(2)
		\Gamma(n)[-\Delta]^{n-m-2}}.\quad
\end{eqnarray}
In the final step, we apply the following  identity  to extract the imaginary parts \cite{Azizi:2017ubq}:
\begin{eqnarray}\label{gamma}
	\Gamma[\frac{D}{2}-n](-\frac{1}{\Delta})^{D/2-n}=\frac{(-1)^{n-1}}{(n-2)!}(-\Delta)^{n-2}ln[-\Delta].
\end{eqnarray}

It is important to note that for dimensions devoid of an imaginary component, their contributions are calculated directly using the standard procedures of the method. Ultimately, the correlation function within the QCD framework is expressed as follows, representing a function of twenty-four distinct Lorentz structures:
\begin{eqnarray}\label{Structures}
	&&\Pi_{\mu}^{\mathrm{QCD}}(p,p',q)=\sum_i\Pi^{\mathrm{QCD}}_{i}(p^{2},p'^{2},q^{2})~ {\mathrm s}_i,
\end{eqnarray}
where $i $ runs from 1 to 24,  and $ {\mathrm S}_i$ denotes the various Lorentz structures presented in Table~\ref{table1}.
\begin{table}[htb]
		{\hfill
		\hbox{\large{
	\begin{tabular}{|c|c|}\hline\hline
		$\large{i}$&$\large{\mathrm{s}_i}$\\
		\hline
		1&$\slashed{p}' \gamma_{\mu}\slashed{p}$\\ \hline
		2&$p_{\mu} \slashed {p}'\slashed {p}$\\ \hline
		3&$p_{\mu}' \slashed {p}'\slashed {p}$\\ \hline	
		4&$p'_\mu\slashed {p}'\gamma_5$\\ \hline
		5&$p'_\mu\slashed {p}'\slashed{p}\gamma_5$\\ \hline
		6&$\slashed {p}'\gamma_\mu\gamma_5$\\ \hline
		7&$\slashed {p}'\gamma_\mu\slashed {p}\gamma_5$\\ \hline
		8&$p_{\mu} \slashed {p}' \slashed{p}\gamma_{5}$\\ \hline 
	\end{tabular}
}}
\hbox{\large{
	\begin{tabular}{|c|c|}\hline\hline
		$\large{i}$&$\large{\mathrm{s}_i}$\\
		\hline
		9&$\slashed{p}' \gamma_{\mu}$\\ \hline
		10&$p_\mu\slashed {p}'\gamma_5$\\ \hline
		11&$p'_\mu\slashed {p}'$\\ \hline	
		12&$p_\mu\slashed {p}'$\\ \hline
		13&$\gamma_\mu\slashed {p}\gamma_5$\\ \hline
		14&$\gamma_{\mu}$\\ \hline
		15&$\gamma_{\mu}\slashed {p}$\\ \hline
		16&$\gamma_{\mu}\gamma_{5}$\\ \hline
		\end{tabular}
}}
	\hbox{\large{
	\begin{tabular}{|c|c|}\hline\hline
		$\large{i}$&$\large{\mathrm{s}_i}$\\
		\hline
		17&$p_\mu\slashed {p}\gamma_5$\\ \hline
		18&$p'_\mu\slashed {p}\gamma_5$\\ \hline
		19&$p'_\mu\slashed {p}$\\ \hline	
		20&$p_\mu\slashed {p}$\\ \hline
		21&$p'_\mu$\\ \hline
		22&$p'_\mu\gamma_5$\\ \hline
		23&$p_\mu$\\ \hline
		24&$p_\mu\gamma_5$\\ \hline 
	\end{tabular}
}}
\hfill}
\caption{Various  Lorentz structures appeared in the calculations.}
\label{table1}
\end{table}
The $\Pi_{i}^{\mathrm{QCD}}(p^2,p'^2,q^2)$ functions  computed on the QCD side are Lorentz invariants, expressed in terms of double dispersion integrals as:
\begin{eqnarray}\label{PiQCD}
	\Pi^{\mathrm{QCD}}_i(p^{2},p'^{2},q^{2})&=&\int_{(2m_c+m_b)^2}^{\infty}ds
	\int_{(2m_c)^2}^{\infty}ds'~\frac{\rho
		^{\mathrm{QCD}}_i(s,s',q^{2})}{(s-p^{2})(s'-p'^{2})} +\Gamma_i(p^2,p'^2,q^2),
\end{eqnarray}
where $\rho_i$ is referred as the spectral density, $\rho_i^{\mathrm{OPE}}(s,s',q^{2})=\frac{1}{\pi}Im\Pi^{OPE}_i(p^2,p'^2,q^2)$.  An alternative  procedure for obtaining  the two-variable spectral density,  $\rho_i(s,s')$,  for a simple term as an example in  the QCD side, and subtraction of higher states and continuum contributions is  provided in the Appendix A (for some details about the  dispersion relation for the two-variable complex function see also Refs.~\cite{Lehmann:1966nxj, Sinha:2020win}.)  $\rho_i^{OPE}$ encompasses the following components:
i) the perturbative part,
ii) the three-mass dimension or the condensation of two quarks, $\langle \overline{q}q\rangle$,
iii) the four-mass dimension or the condensation of two gluons, $\langle G^2\rangle$.  $\Gamma_i$ represents the contribution of the five-mass dimension, which is calculated directly. The  $\rho_i$ and $\Gamma_i$ for the structure $\gamma_\mu \gamma_{5}$,  as an example,  are presented in the Appendix B.  We present the results for this structure related to the form factor $ G_1 $ in the Appendix B, since this structure revives nonzero expressions for all the perturbative and nonperturbative contributions in momentum space. By adopting the  quark-hadron duality assumption, continuum thresholds, denoted as $s_0$ and $s'_0$, emerge in both the initial and final states. Ultimately, by applying double Borel transformation to suppress the contributions coming from the excited and continuous states, the form of the correlation function utilized in the QCD side is obtained as follows:
\begin{eqnarray}\label{qcd part2}
	&&\Pi^{\mathrm{QCD}}_i (M^2,M'^2,s_0,s'_0,q^2)=\int _{(2m_c+m_b)^2}^{s_0} ds\int _{(2m_c)^2}^{s'_0}ds' e^{-s/M^2} e^{-s'/M'^2}\rho
	^{\mathrm{QCD}}_{i}(s,s',q^{2})+\tilde{\Gamma}_i(M^2,M'^2,q^2),
\end{eqnarray}
where $ \tilde{\Gamma}_i(M^2,M'^2,q^2) $  is the double Borel transformed form of $ \Gamma_i(p^2,p'^2,q^2)$.  Note that, the contribution of five-mass dimensional nonperturbative operator  vanishes after applying the  double Borel transformation for the  structure $\gamma_\mu \gamma_{5}$.
To ultimately derive the six form factors,    various corresponding coefficients of structures from both the hadronic and QCD sides are matched.	
\section {numerical analysis }\label{Sec:three}
\begin{figure}[h!] 
	\includegraphics[totalheight=4.8cm,width=4.9cm]{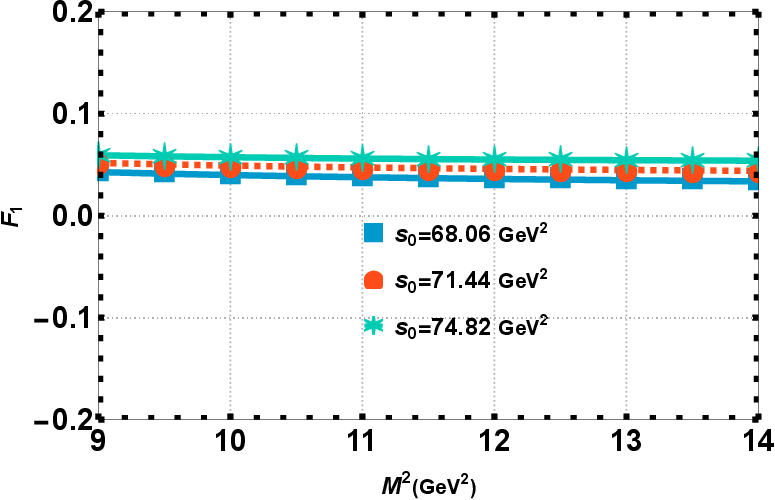}
	\includegraphics[totalheight=4.8cm,width=4.9cm]{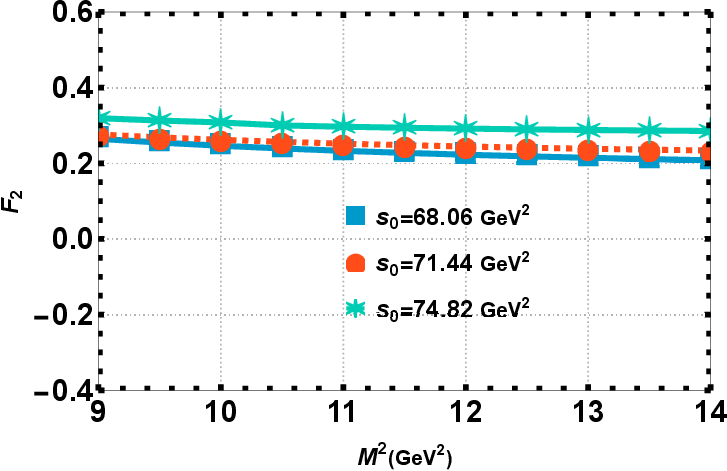}
	\includegraphics[totalheight=4.8cm,width=4.9cm]{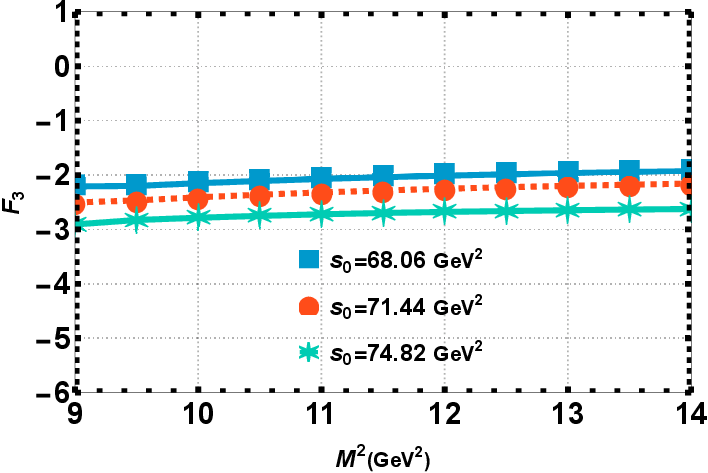}
	\includegraphics[totalheight=4.8cm,width=4.9cm]{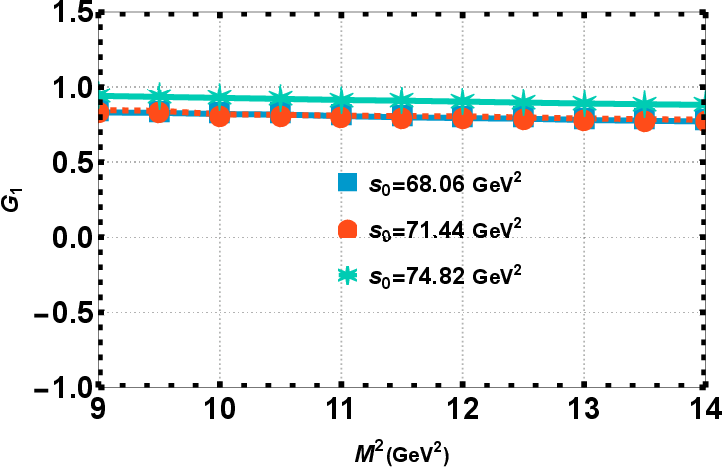}
	\includegraphics[totalheight=4.8cm,width=4.9cm]{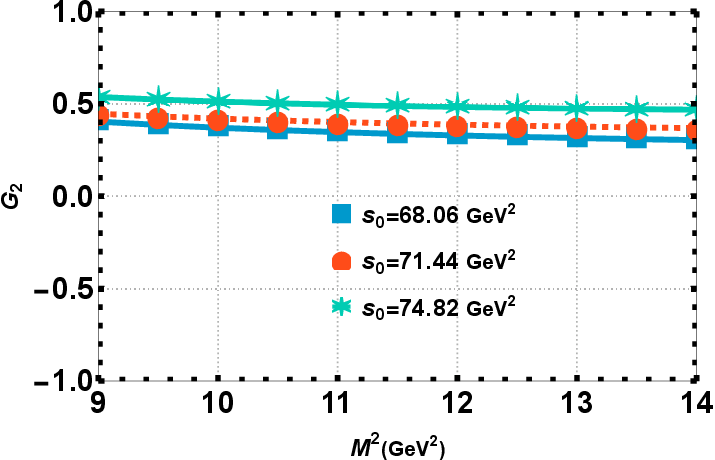}
	\includegraphics[totalheight=4.8cm,width=4.9cm]{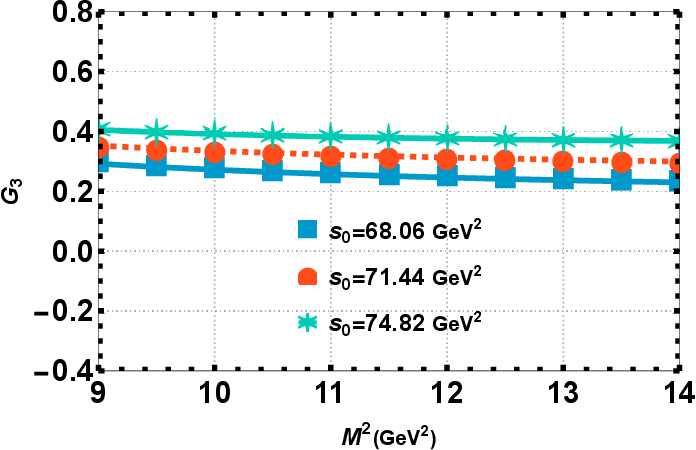}
	\caption{Form factors as functions of the Borel parameter $M^2$ at  various values 
		of the parameter $s_0$,   $q^2=0$ and average values of other auxiliary parameters ($s'_0=15.40~\mathrm{GeV}^{2}$, $M'^2=5.00 ~\mathrm{GeV}^{2}$ and  $\cos\theta =0.60$).}\label{Fig:BorelM}
\end{figure}

\begin{figure}[h!]
	\includegraphics[totalheight=4.8cm,width=4.9cm]{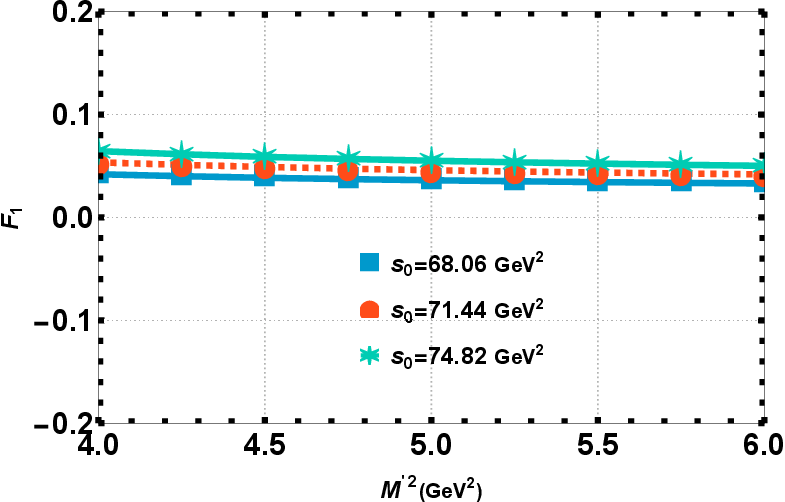}
	\includegraphics[totalheight=4.8cm,width=4.9cm]{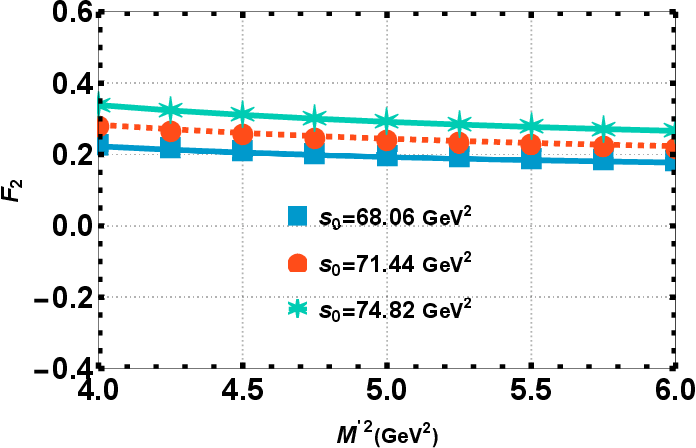}
	\includegraphics[totalheight=4.8cm,width=4.9cm]{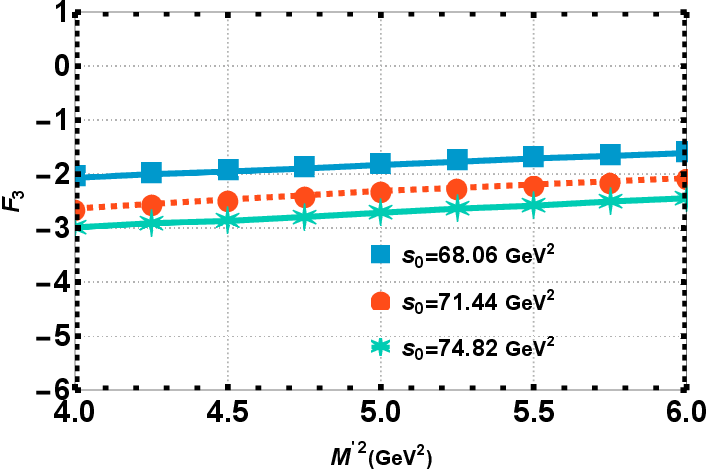}
	\includegraphics[totalheight=4.8cm,width=4.9cm]{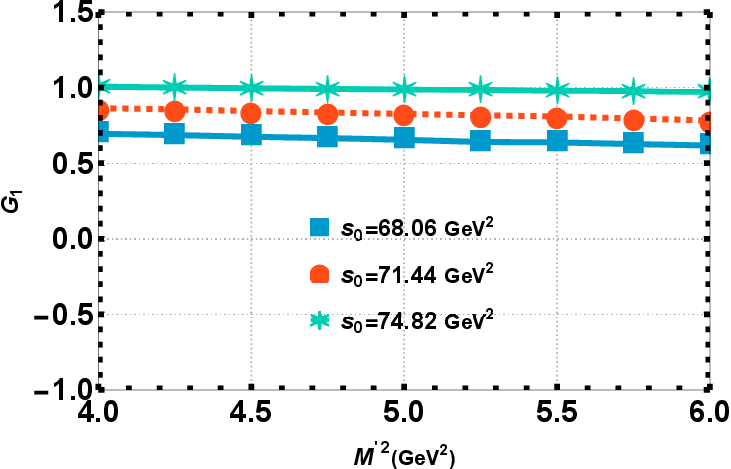}
	\includegraphics[totalheight=4.8cm,width=4.9cm]{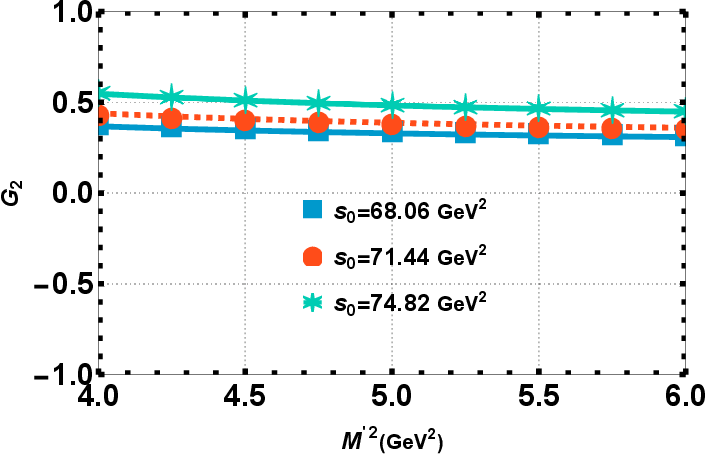}
	\includegraphics[totalheight=4.8cm,width=4.9cm]{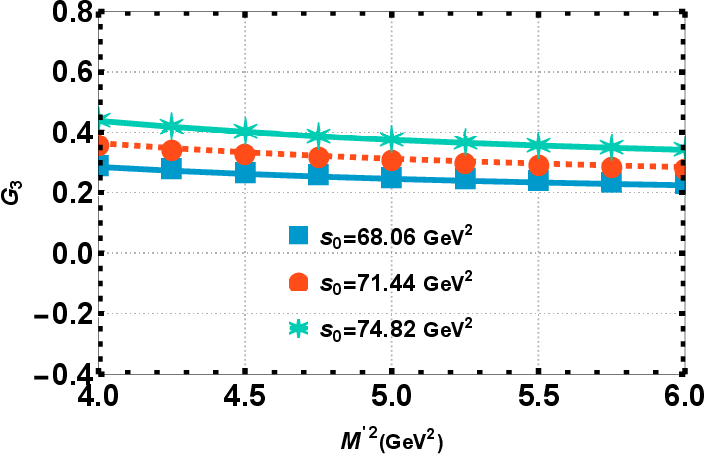}
	\caption{Form factors as functions of the Borel parameter $M'^2$ at various values 
		of the parameter $s_0$,  $q^2=0$ and average values of other auxiliary parameters ($s'_0=15.40~\mathrm{GeV}^{2}$, $M^2=11.50~\mathrm{GeV}^{2}$ and  $\cos\theta =0.60$).} \label{Fig:BorelMM}
\end{figure}

\begin{figure}[h!] 
	\includegraphics[totalheight=4.8cm,width=4.9cm]{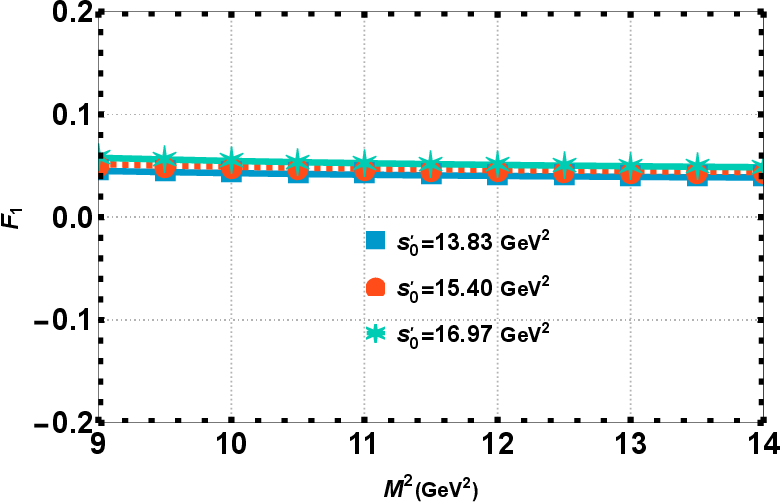}
	\includegraphics[totalheight=4.8cm,width=4.9cm]{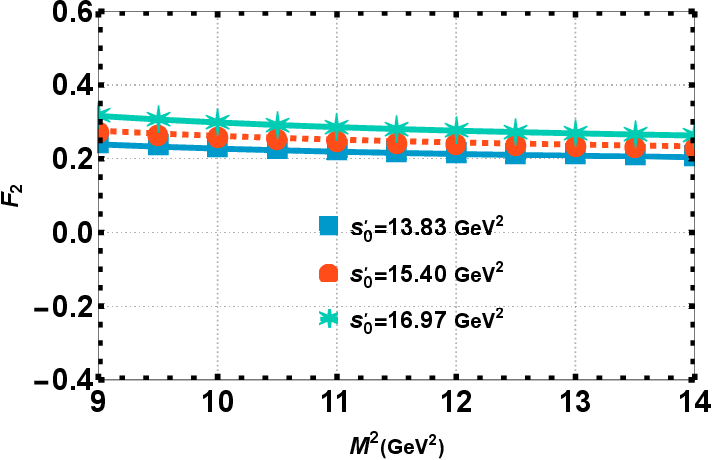}
	\includegraphics[totalheight=4.8cm,width=4.9cm]{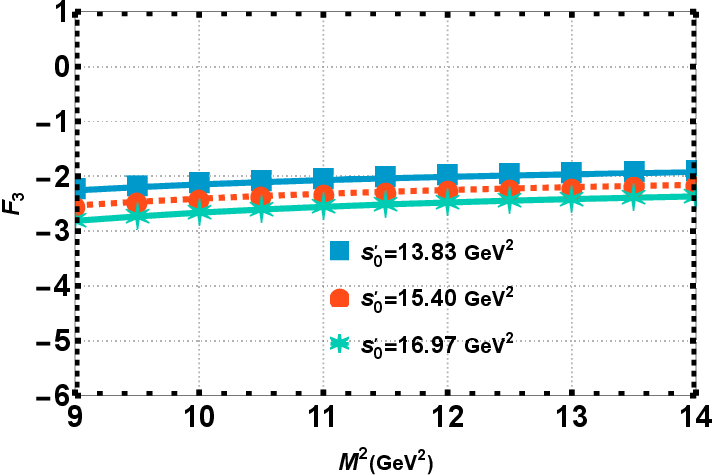}
	\includegraphics[totalheight=4.8cm,width=4.9cm]{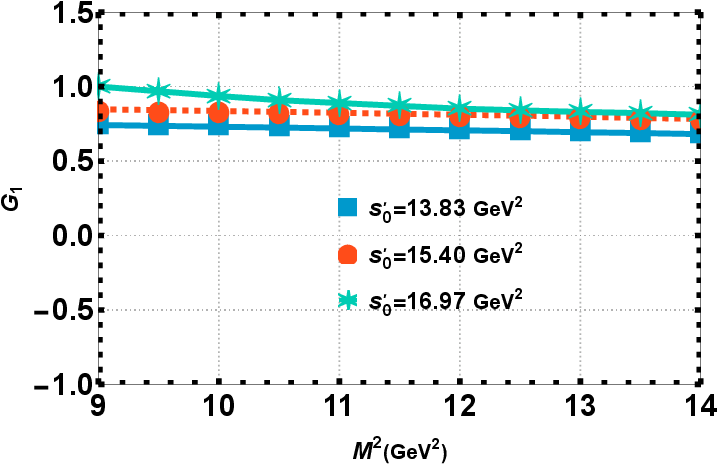}
	\includegraphics[totalheight=4.8cm,width=4.9cm]{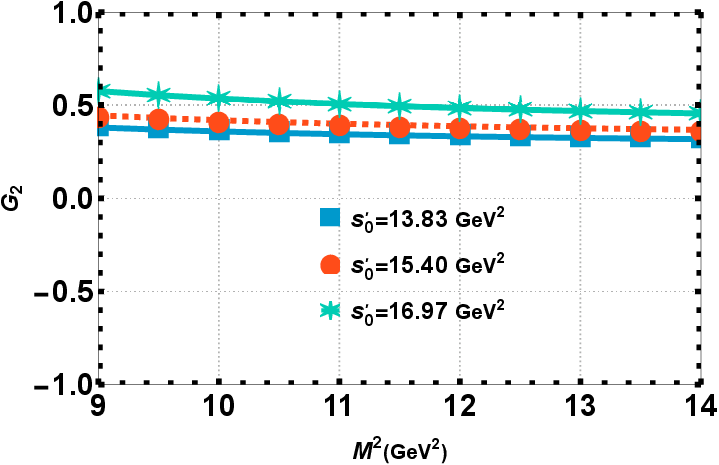}
	\includegraphics[totalheight=4.8cm,width=4.9cm]{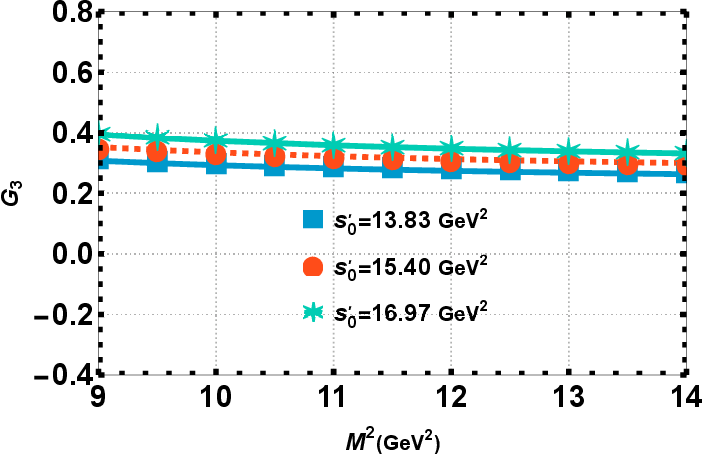}
	\caption{Form factors as functions of the Borel parameter $M^2$ at  various values 
		of the parameter $s'_0$,   $q^2=0$ and average values of other auxiliary parameters($s_0=71.44~\mathrm{GeV}^{2}$, $M'^2=5.00~\mathrm{GeV}^{2}$ and  $\cos\theta =0.60$).}\label{Fig:BorelMs'}
\end{figure}

\begin{figure}[h!]
	\includegraphics[totalheight=4.8cm,width=4.9cm]{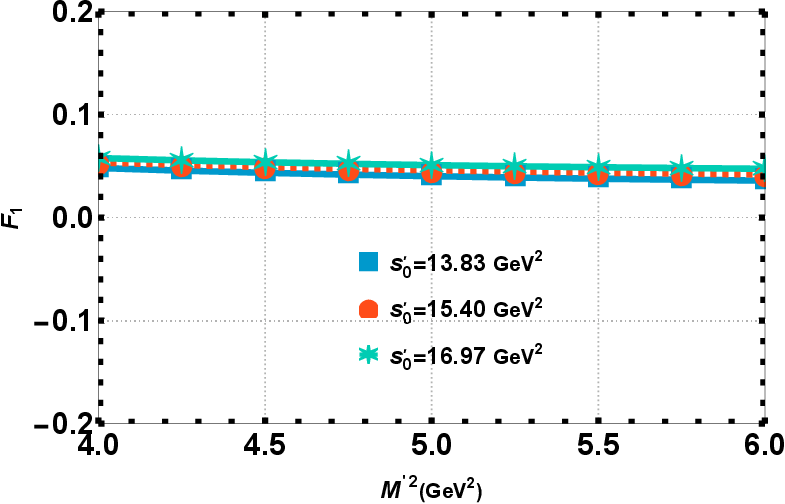}
	\includegraphics[totalheight=4.8cm,width=4.9cm]{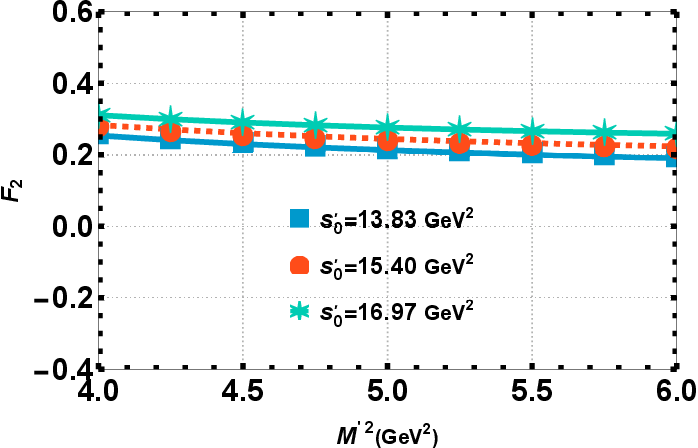}
	\includegraphics[totalheight=4.8cm,width=4.9cm]{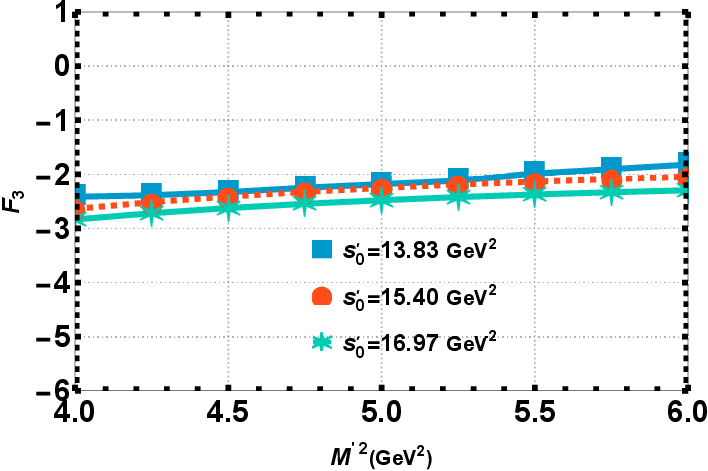}
	\includegraphics[totalheight=4.8cm,width=4.9cm]{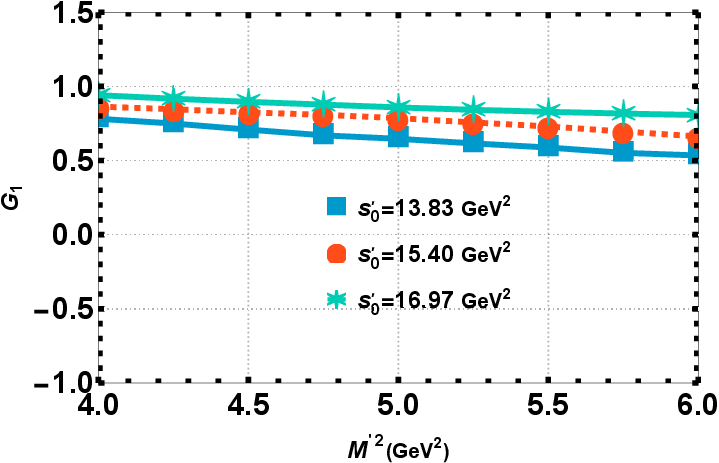}
	\includegraphics[totalheight=4.8cm,width=4.9cm]{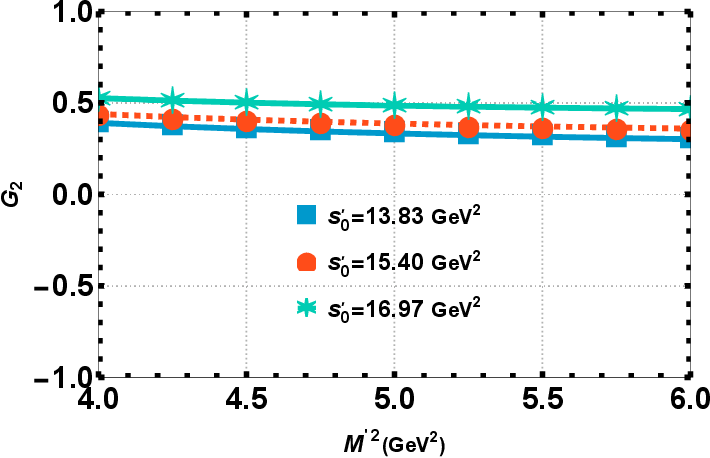}
	\includegraphics[totalheight=4.8cm,width=4.9cm]{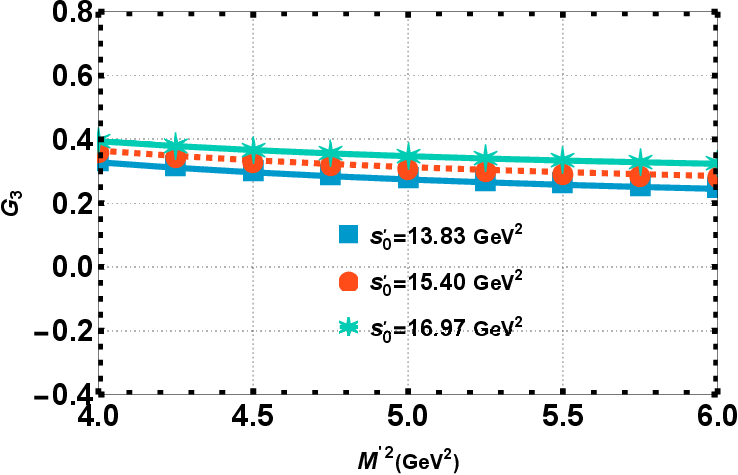}
	\caption{Form factors as functions of the Borel parameter $M'^2$ at various values 
		of the parameter $s'_0$,  $q^2=0$ and average values of other auxiliary parameters ($s'_0=71.44~\mathrm{GeV}^{2}$, $M^2=11.50 ~\mathrm{GeV}^{2}$ and  $\cos\theta =0.60$).} \label{Fig:BorelMMs'}
\end{figure}

In this section, we present the numerical results for the form factors defining   the semileptonic decay $ \Omega_{ccb}^{+}\rightarrow \Xi^{++}_{cc}  ~{\ell}\bar\nu_{\ell}$. A comprehensive set of input parameters utilized in the calculations is depicted in Table~\ref{inputParameter}.
\begin{table}[h!]
		\begin{tabular}{|c|c|}
		\hline 
		Parameters                                             &  Values  \\
		\hline \hline
		$ m_b$                                                 & $(4.18^{+0.03}_{-0.02})~ \mathrm{GeV}$ \cite{ParticleDataGroup:2022pth}\\
		$ m_c$                                                 & $(1.27\pm0.02)~ \mathrm{GeV}$ \cite{ParticleDataGroup:2022pth}\\
			$ m_e $                                                & $ 0.51~\mathrm{MeV}$ \cite{ParticleDataGroup:2022pth}\\
		$ m_\mu $                                              & $ 105~\mathrm{MeV}$ \cite{ParticleDataGroup:2022pth}\\
		$ m_\tau $                                             & $ 1.776~\mathrm{GeV}$ \cite{ParticleDataGroup:2022pth}\\
		$ m_{\Omega_{ccb}^{+}}$                                       & $ (8.15{}^{0.27}_{0.23}) ~\mathrm{GeV}$ \cite{Najjar:2024deh}\\
		$ m_{\Xi^{++}_{cc}} $                                      & $ (3621.6\pm0.4)~\mathrm{MeV}$  \cite{ParticleDataGroup:2022pth} \\
		$ G_{F} $                                              & $ 1.17\times 10^{-5}~ \mathrm{GeV^{-2}}$ \cite{ParticleDataGroup:2022pth}\\
		$ V_{ub} $                                             & $ (3.82\pm0.20)\times 10^{-3} $  \cite{ParticleDataGroup:2022pth}\\
		$ m^2_0 $                                              & $ (0.8\pm0.2)~ \mathrm{GeV^2}$ \cite{Belyaev:1982sa,Belyaev:1982cd,Ioffe:2005ym} \\
		$\langle \bar{u} u\rangle$         & $-(0.24\pm0.01)^3 ~\mathrm{GeV^3}$  \cite{Belyaev:1982sa,Belyaev:1982cd} \\
		$\langle \frac{\alpha_s}{\pi} G^2 \rangle $ & $(0.012\pm0.004)$ $~\mathrm{GeV}^4 $ \cite{Belyaev:1982sa,Belyaev:1982cd,Ioffe:2005ym}\\
		$\lambda_{\Omega_{ccb}^{+}}$                               &$0.44{}^{0.23}_{0.25}~  \mathrm{GeV^3}$  \cite{Najjar:2024deh}\\
		$\lambda_{\Xi^{++}_{cc}}$                               & $0.16\pm0.04~  \mathrm{GeV^3}$ \cite{ShekariTousi:2024mso}\\
		\hline
	\end{tabular}
\caption{Input parameters used in calculations.}\label{inputParameter}
\end{table}
 In addition to the parameters listed in Table~\ref{inputParameter}, five auxiliary parameters are introduced, which were referenced throughout the calculations: the Borel parameters $M^2$ and $M'^2$, the continuum thresholds $s_0$ and $s'_0$, as well as the arbitrary mixing parameter $\beta$. The physical quantities must exhibit minimal dependence on variations in these parameters. Within the framework of  QCD sum rule method, to ascertain the operational window for the aforementioned auxiliary parameters, two conditions must be satisfied:  the pole contribution(PC) dominance and convergence of the OPE series. These conditions can be imposed  as follows:	
\begin{equation} \label{PC}
	PC=\frac{\Pi^{QCD}(M^2,M'^2,s_0,s'_0)}{\Pi^{QCD}(M^2,M'^2,{\infty},{\infty})}\geq0.5,
\end{equation}
and	
\begin{equation} \label{PC2}
	R(M^2, M'^2)=\frac{\Pi^{^{QCD}(dim4+dim5)}(M^2,M'^2,s_0,s'_0)}{\Pi^{QCD}(M^2,M'^2,s_0,s'_0)}\leq0.05.
\end{equation}	
	The PC criterion, as articulated in Eq.~(\ref{PC}),  requires that the PC exceeds the  contributions coming  from the  higher and continuum states, and help us determine  the upper limits for the Borel parameters. The convergence of the OPE series, as expressed in Eq.~(\ref{PC2}), establishes the lower limits for these parameters. Consequently, based on these criteria,  the Borel windows are implemented as follows:
\begin{eqnarray}
	&&9~\mathrm{GeV^2}\leq M^2 \leq 14~\mathrm{GeV^2}, \notag\\
	\mbox{and} \notag\\
	&&4~\mathrm{GeV^2} \leq M'^2 \leq 6~\mathrm{GeV^2}.
\end{eqnarray}
The parameters $s_0$ and $s'_0$ are auxiliary variables whose values are not entirely optional; their values depend on the energy  of the excited states in both the initial and final channels. Although there is a lack of experimental data regarding the excited states of triply and doubly heavy baryons, our analysis affirm the optimal stability of the sum rules within the following intervals:	
\begin{eqnarray}
	&&(m_{\Omega_{ccb}^{+}}+0.1)^2~ \mathrm{GeV^2} \leq s_{0} \leq (m_{\Omega_{ccb}^{+}}+0.5)^2~ \mathrm{GeV^2},\notag\\
	\mbox{and} \notag\\
	&&(m_{\Xi^{++}_{cc}}+0.1)^2~\mathrm{GeV^2}\leq s'_{0} \leq (m_{\Xi^{++}_{cc}}+0.5)^2~ \mathrm{GeV^2}.
\end{eqnarray}	
\begin{figure}[h!]
	\includegraphics[totalheight=6cm,width=8cm]{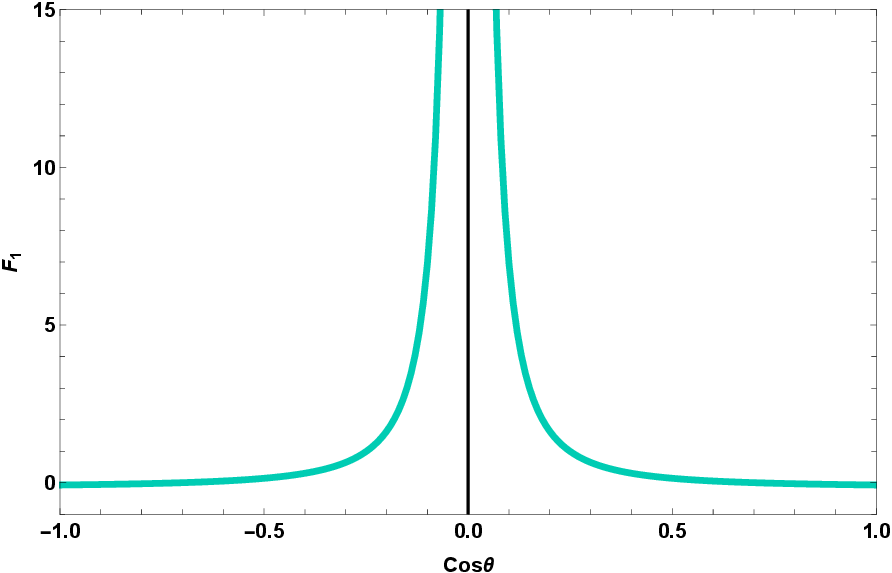}
	\caption{Variations of $F_1$ form factor corresponding to the structure $\slashed {p}'\gamma_{\mu} $ with respect to  $\cos\theta$ at  the central values of  $s_0$, $s'_0$, $M^2$ and $M'^2$ and at $q^2=0$.} \label{Fig:beta}
\end{figure}
In addition to the Borel parameters and continuum
thresholds,  the other auxiliary parameter
to be set is $\beta$ entered  Eqs. ~( \ref{cur1}) and 
	(\ref{eq:CorrF2} ). While $\beta$ can be chosen to be different for the two interpolating currents in initial and final channels,  our analyses show that they belong to working regions that have a major overlap.  The optimization of the physical quantities with respect to these parameters lead us to take this parameter the same for two interpolating currents.  The range of the auxiliary mathematical parameter  $\beta$ can be evaluated as follows: since this parameter spans the entire interval from $- \infty $ to $+\infty$, we introduce a new parameter defined as $\beta=\tan\theta$ to incorporate all possible values for $\beta$ such that $	-1.0 \le\ \cos\theta \le\ 1.0$. Fig~\ref{Fig:beta} exemplifies the variation of $F_1$ as a function of $\cos\theta$, allowing for the identification of the operational region for $\cos\theta$. Based on numerical analysis and insights  from  Fig~\ref{Fig:beta}, we conclude that the optimal regions conducive to the stability of form factors with respect to variations in $\cos\theta$  are delineated as follows:	
\begin{eqnarray}\label{beta fun}
-0.7 \le\ \cos\theta \le\ -0.5~~~~~~~	\mbox{and} ~~~~~~  0.5 \le\ \cos\theta \le\ 0.7.
\end{eqnarray}	
	To demonstrate the stability of the  six form factors with respect to the auxiliary parameters,  Figs.~\ref{Fig:BorelM}, \ref{Fig:BorelMM}, \ref{Fig:BorelMs'} and \ref{Fig:BorelMMs'} are provided. In these figures, $F_1$, $F_2$, $F_3$, $G_1$, $G_2$, and $G_3$ correspond to the structures denoted as $\slashed {p}'\gamma_{\mu}$, $p_\mu \slashed {p}'$, $p'_{\mu}\slashed {p}$, $\slashed {p}'\gamma_\mu \slashed {p} \gamma_5$, $p_\mu \slashed {p}' \gamma_{5}$ and  $p_\mu \gamma_{5}$, respectively.  In Figs.~\ref{Fig:BorelM} and \ref{Fig:BorelMM} the variations of form factors are depicted as functions of $M^2$ and $M'^2$ respectively, at three fixed values of  $s_0$   along with other parameters at their average values. In Figs.~\ref{Fig:BorelMs'} and \ref{Fig:BorelMMs'} the variations of form factors are depicted as functions of $M^2$ and $M'^2$ respectively, at three fixed values of  $s'_0$ while other parameters are kept at their average values.  It is important to note that all graphs are generated at the average value of $\cos\theta$ with $q^2=0$. We see that the form factors demonstrate good stability with respect to the auxiliary parameters in their working intervals.

\begin{figure}[h!] 
	\includegraphics[totalheight=4.5cm,width=4.8cm]{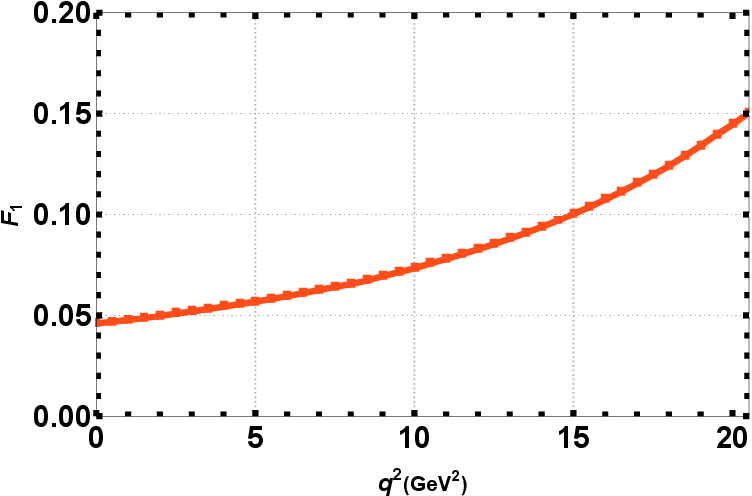}
	\includegraphics[totalheight=4.5cm,width=4.8cm]{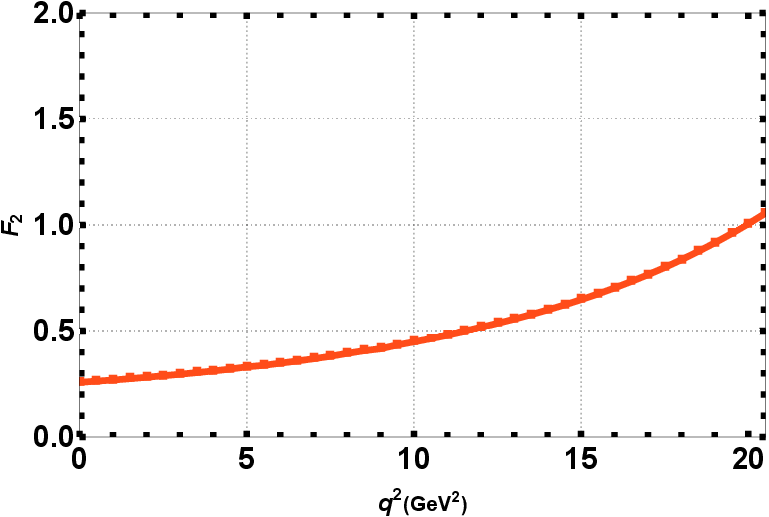}
	\includegraphics[totalheight=4.5cm,width=4.8cm]{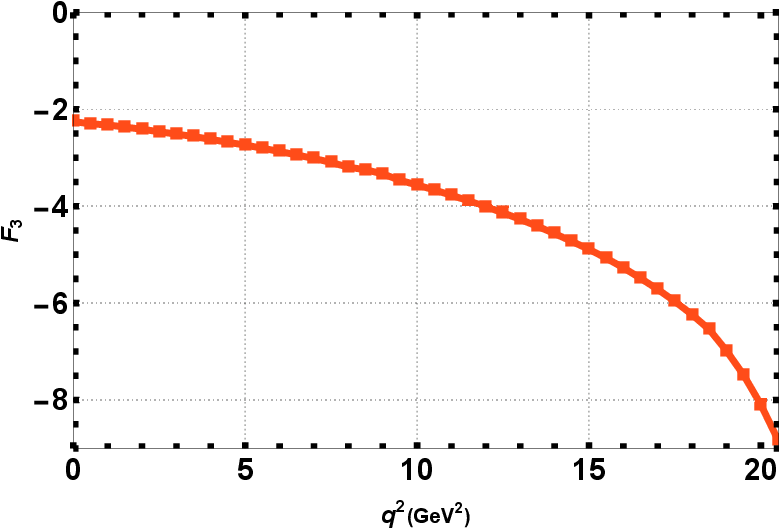}
	\includegraphics[totalheight=4.5cm,width=4.8cm]{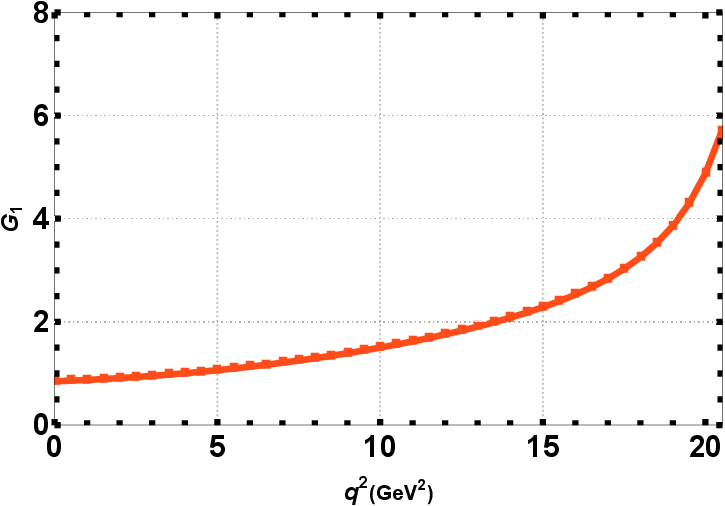}
	\includegraphics[totalheight=4.5cm,width=4.8cm]{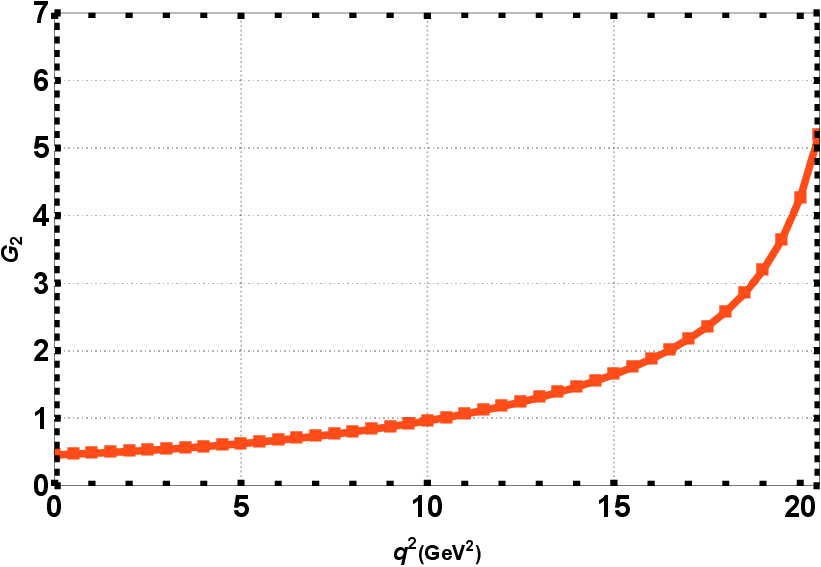}
	\includegraphics[totalheight=4.5cm,width=4.8cm]{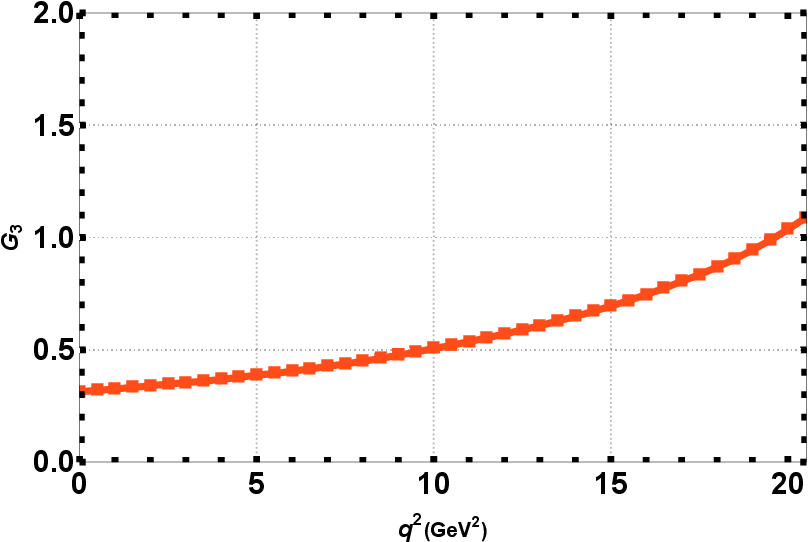}
	\caption{The form factors $F_1$,  $F_2$, $F_3$,  $G_1$ , $G_2$ and $G_3$, corresponding to the structures in Table \ref{Tab:parameterfit}, as functions of $q^2$ at average values of auxiliary parameters.}\label{Fig:formfactor1}
\end{figure}

\begin{figure}[h!] 
	\includegraphics[totalheight=4.5cm,width=4.8cm]{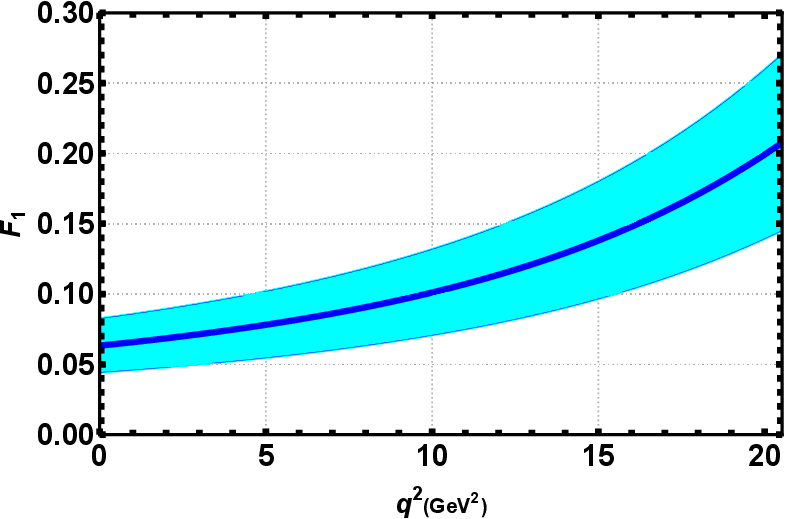}
	\includegraphics[totalheight=4.5cm,width=4.8cm]{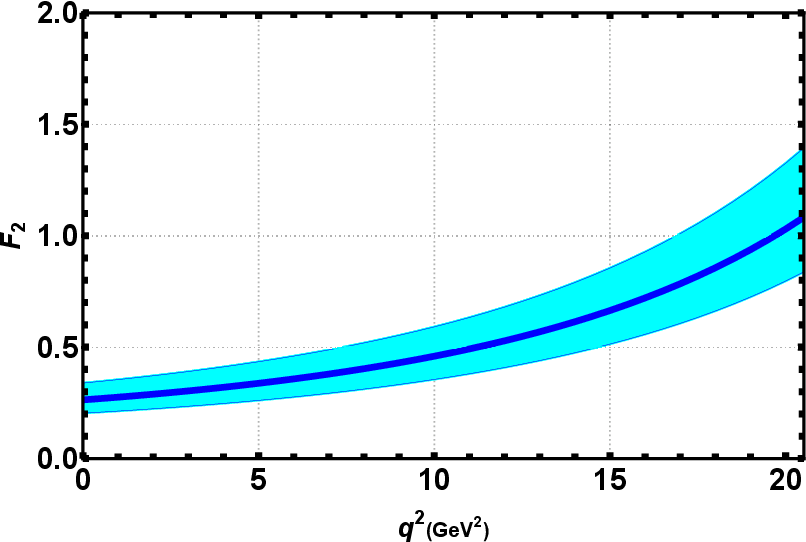}
	\includegraphics[totalheight=4.5cm,width=4.8cm]{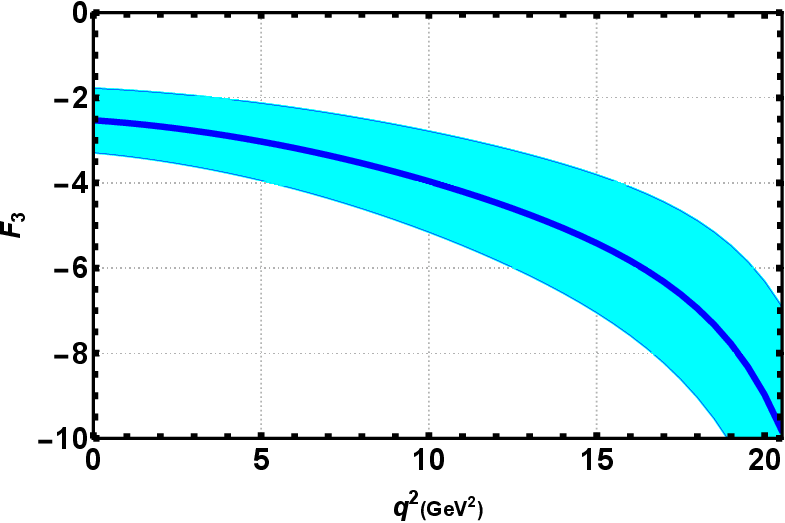}
	\includegraphics[totalheight=4.5cm,width=4.8cm]{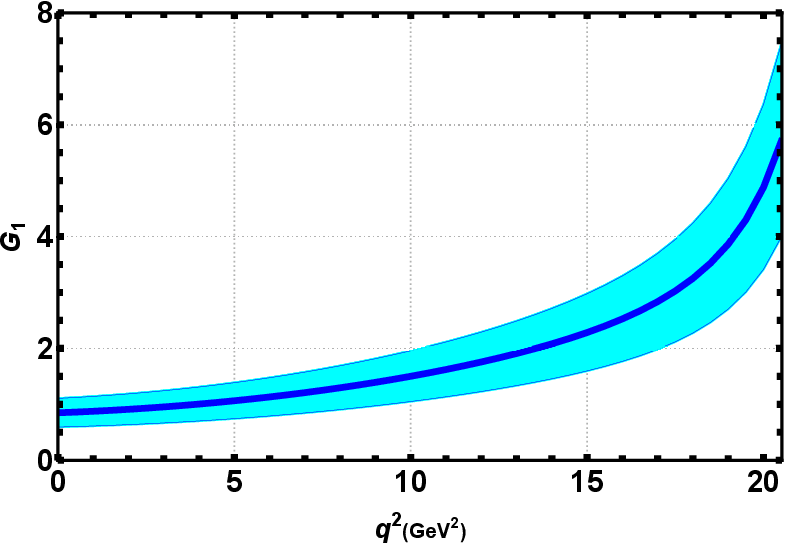}
	\includegraphics[totalheight=4.5cm,width=4.8cm]{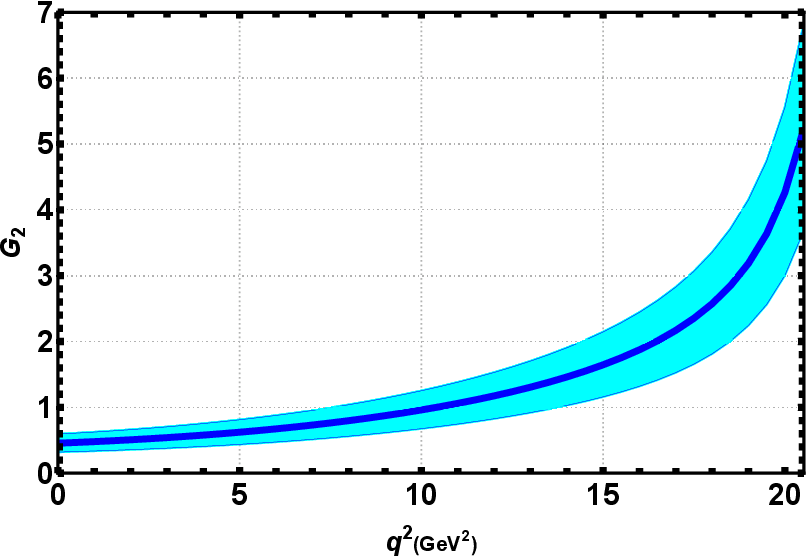}
	\includegraphics[totalheight=4.5cm,width=4.8cm]{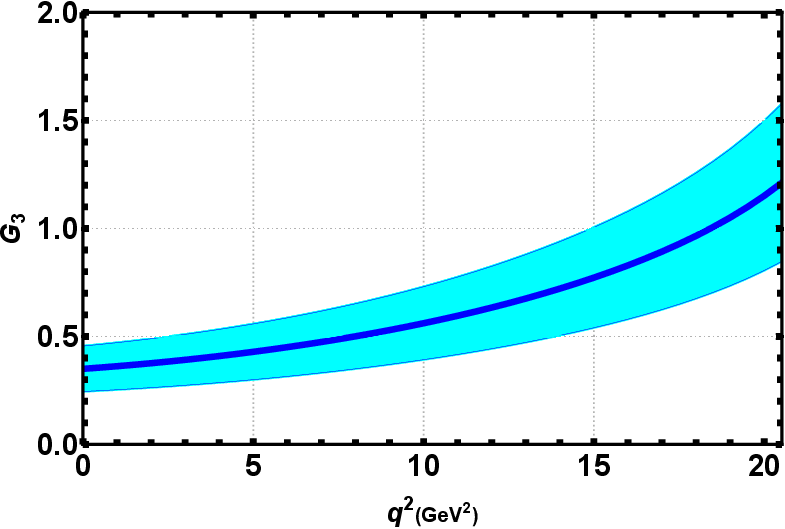}
	\caption{The form factors corresponding to the structures in Table \ref{Tab:parameterfit} with their errors at the average values of auxiliary parameters.}\label{Fig:formfactorserror1}
\end{figure}

Following the identification of the operational windows for all the auxiliary parameters, we  proceed to examine the behavior of the form factors as  functions of $q^2$. Our analysis indicates that the fitting of the form factors is effectively achieved using the following expression:
\begin{equation} \label{fitffunction}
	{\cal F}(q^2)=\frac{{\cal
			F}(0)}{\displaystyle\left(1-a_1\frac{q^2}{m^2_{\Omega_{ccb}^{+}}}+a_2
		(\frac{q^2}{m_{\Omega_{ccb}^{+}}^2})^2+a_3(\frac{q^2}{m_{\Omega_{ccb}^{+}}^2})^3+a_4(\frac{q^2}{m_{\Omega_{ccb}^{+}}^2})^4\right)}.
\end{equation}
In this context, the parameters  , ${\cal F}(0)$, $a_1$, $a_2$, $a_3$ and $a_4$ serve as the fitting parameters corresponding to the structures delineated in Table~\ref{Tab:parameterfit}. It is important to note that these quantities are computed in the average values of the auxiliary parameters.  It is noteworthy that the uncertainty associated with the form factors presented in Table~\ref{Tab:parameterfit} arises from both of the auxiliary parameters and the input ones.  It should be noted that the structures in Table~\ref{Tab:parameterfit} are not unique to find the form factors as is clear from the presented results in both the physical and QCD sides. In the framework of sum rule method,  the form factors are practically obtained to be structure-dependent.  Generally,  structures with more momenta lead to more stability with respect to the variations of the auxiliary parameters.  In this study,  the selected structures are based on the operational domains of the auxiliary parameters, wherein the uncertainty of the results is relatively minimal and requirements of the method, discussed above,  are better satisfied. The variations of the form factors $F_1$, $F_2$, $F_3$, $G_1$, $G_2$, and $G_3$  as functions of $q^2$ within the permissible range $ m_l^2\leq q^2 \leq (m_{\Omega_{ccb}^{+}}-m_{\Xi^{++}_{cc}})^2$  for the structures  $\slashed {p}'\gamma_{\mu}$, $p_\mu \slashed {p}'$, $p'_{\mu}\slashed {p}$, $\slashed {p}'\gamma_\mu \slashed {p} \gamma_5$, $p_\mu \slashed {p}' \gamma_{5}$ and  $p_\mu \gamma_{5}$, respectively,  are illustrated in Fig.~\ref{Fig:formfactor1}. Fig.~\ref{Fig:formfactorserror1} depicts these variations, incorporating uncertainties in the calculation of form factors. In the subsequent section, we will employ the fitting functions of the form factors as primary input parameters to evaluate the  decay widths at different lepton channels.
\begin{table}[h!]
	\begin{ruledtabular}
		\begin{tabular}{|c|c|c|c|c|c|c|}
			& $F_1(q^2): \slashed {p}'\gamma_{\mu} $ & $F_2(q^2):p_{\mu}\slashed {p'}$  & $F_3(q^2):p'_{\mu}\slashed {p}$   & $G_1(q^2):\slashed {p}'\gamma_{\mu} \slashed{p}\gamma_5 $ & $G_2(q^2):p_{\mu}\slashed {p}'\gamma_5$  & $G_3(q^2):p_{\mu}\gamma_5$       \\
			\hline
			${\cal F}(q^2=0)$ & $0.06\pm0.01$        & $0.26{}^{+0.07}_{-0.05}$      & $-2.52{}^{+0.74}_{-0.75}$     & $0.84\pm0.25$  & $0.45\pm0.13$  & $0.34\pm0.10$  \\
			$a_1$           & $2.52$          & $2.92$            &$1.41$            & $1.80$           & $3.12$            &$ 2.14$              \\
			$a_2$           & $-0.25$         & $-0.82$           &$-15.54$              & $-17.87$          & $-10.14$           & $-5.94$           \\
			$a_3$           & $3.43$            & $11.13$           &$78.45$           & $92.95$          & $67.61$           & $29.93$           \\
			$a_4$           & $0.99$           & $-11.32$           &$-124.53$             & $-145.86$         & $-106.61$          & $-39.92$           \\
		\end{tabular}
		\caption{Parameters of the fit functions for different form factors corresponding to $\Omega_{ccb}^{+}\to\Xi^{++}_{cc} l\bar\nu_{\ell}$ transition.}\label{Tab:parameterfit}
	\end{ruledtabular}
\end{table}

\section {decay width }\label{Sec:four}
In this section, the decay widths of  semileptonic mode $ \Omega_{ccb}^{+}\rightarrow \Xi^{++}_{cc}  ~{\ell}\bar\nu_{\ell}$ across all lepton channels are computed utilizing the fitted form factors derived in the preceding section. The differential decay width is given by: 
\begin{equation}\label{eq:dgamma}
	\frac{d\Gamma(\Omega_{ccb}^{+}\rightarrow \Xi^{++}_{cc}  ~{\ell}\bar\nu_{\ell})}{dq^2}=\frac{G_F^2}{(2\pi)^3}
	|V_{ub}|^2\frac{\lambda^{1/2}(q^2-m_\ell^2)^2}{48m_{\Omega_{ccb}^{+}}^3q^2}{\cal
		H}_{tot}(q^2),
\end{equation}
 where $\lambda\equiv\lambda(m^2_{\Omega_{ccb}^{+}}, m^2_{\Xi^{++}_{cc}}, q^2)=m^4_{\Omega_{ccb}^{+}}+m^4_{\Xi^{++}_{cc}}+q^4-2(m^2_{\Omega_{ccb}^{+}}m^2_{\Xi_{cc}}+m^2_{\Omega_{ccb}^{+}}q^2+m^2_{\Xi_{cc}}q^2)$, and $m_l$ represents the lepton mass $(e  ,\mu, \tau)$.  ${\cal
 	H}_{tot}(q^2)$ denotes total helicity, which is defined as: 
 \begin{equation}
 	\label{eq:hh}
 	{\cal H}_{tot}(q^2)=[{\cal H}_U(q^2)+{\cal H}_L(q^2)] \left(1+\frac{m_\ell^2}{2q^2}\right)+\frac{3m_\ell^2}{2q^2}{\cal H}_S(q^2).
 \end{equation} 
The relevant parity-conserving helicity structures are expressed in terms of the total
helicity amplitudes as follows:
\begin{eqnarray}
	\label{eq:hhc}
	&&{\cal H}_U(q^2)=|H_{+1/2,+1}|^2+|H_{-1/2,-1}|^2,\notag\\
	&&{\cal H}_L(q^2)=|H_{+1/2,0}|^2+|H_{-1/2,0}|^2,\notag\\
	&&{\cal H}_S(q^2)=|H_{+1/2,t}|^2+|H_{-1/2,t}|^2,
\end{eqnarray}
In this set of expressions, different helicity amplitude are expressed in terms of the transition $F_i$ and $G_i$ form factors:
\begin{eqnarray}
	\label{eq:ha}
	H^{V,A}_{+1/2,\, 0}&=&\frac1{\sqrt{q^2}}{\sqrt{2m_{\Omega_{ccb}^{+}}m_{\Xi_{cc}^{++}}(\sigma\mp 1)}}
	[(m_{\Omega^{+}_{ccb}} \pm m_{\Xi_{cc}^{++}}){\cal F}^{V,A}_1(\sigma) \pm m_{\Xi_{cc}^{++}}
	(\sigma\pm 1){\cal F}^{V,A}_2(\sigma)\cr
	&& \pm m_{\Omega^{+}_{ccb}} (\sigma\pm 1){\cal F}^{V,A}_3(\sigma)],\cr
	H^{V,A}_{+1/2,\, 1}&=&-2\sqrt{m_{\Omega^{+}_{ccb}}m_{\Xi_{cc}^{++}}(\sigma\mp 1)}
	{\cal F}^{V,A}_1(\sigma),\cr
	H^{V,A}_{+1/2,\, t}&=&\frac1{\sqrt{q^2}}{\sqrt{2m_{\Omega^{+}_{ccb}}m_{\Xi_{cc}^{++}}(\sigma\pm 1)}}
	[(m_{\Omega^{+}_{ccb}} \mp m_{\Xi_{cc}^{++}}){\cal F}^{V,A}_1(\sigma) \pm(m_{\Omega^{+}_{ccb}}- m_{\Xi_{cc}^{++}} \sigma
	){\cal F}^{V,A}_2(\sigma)\cr
	&& \pm (m_{\Omega^{+}_{ccb}} \sigma- m_{\Xi_{cc}^{++}}){\cal F}^{V,A}_3(\sigma)],
\end{eqnarray}
where
\begin{equation}
\sigma=\frac{m_{\Omega^{+}_{ccb}}^2+m_{\Xi_{cc}^{++}}^2-q^2}
{2m_{\Omega^{+}_{ccb}}m_{\Xi_{cc}^{++}}}.
\end{equation}
In this context,  ${\cal F}^V_i\equiv F_i$ and  ${\cal F}^A_i\equiv G_i$ ($i=1,2,3$) illustrate vector and axial form factors. The upper (lower) sign is associated with the vector (axial) contribution. $H^{V,A}_{h',\,h_W}$ are utilized to represent  the helicity amplitudes for weak decays including the vector (V) and the axial vector (A) currents, and $h'$ and $h_W$ indices stand for  the helicities of the final baryon and the virtual W-boson, respectively. The amplitudes for negative helicity values can be derived using the relation:
\begin{equation}
	H^{V,A}_{-h',\,-h_W}=\pm H^{V,A}_{h',\,h_W}.
\end{equation}
The total helicity amplitude for the V-A current is thus expressed by:
\begin{equation}
H_{h',\,h_W}=H^{V}_{h',\,h_W}-H^{A}_{h',\,h_W}.
\end{equation}

The calculated decay widths for all lepton channels, along with their associated errors, are provided in Table~\ref{DECAY}. Due to the lack of experimental data on the lifetimes of triply heavy baryons, it is currently not possible to compute the branching fractions. However, the decay width itself serves as an important indicator for guiding experimental searches and theoretical analysis. The decay width is expected to show variations based on the choice of lepton channels, primarily due to the mass differences among the leptons involved. Specifically, as the lepton mass increases (moving from $e$ or $\mu$  to $\tau$), the decay width tends to decrease. This trend is consistent with the fact that decays involving heavier leptons are kinematically suppressed. These findings are crucial in providing a benchmark for upcoming experiments, as they enable experimentalists to estimate decay probabilities and search for corresponding signals.
\begin{table}[h!]
	\begin{ruledtabular}
		\begin{tabular}{|c|c|c|}
			 $\Gamma~[\Omega^{+}_{ccb}\to\Xi_{cc}^{++} e~ {\overline{\nu}}_{e}]\times10^{14}$ & $\Gamma~[\Omega^{+}_{ccb}\to\Xi_{cc}^{++} \mu~ {\overline{\nu}}_{\mu}]\times10^{14}$ &$\Gamma~[\Omega^{+}_{ccb}\to\Xi_{cc}^{++} \tau~ {\overline{\nu}}_{\tau}]\times10^{15}$   \\   
			\hline
		$1.23{}^{+0.28}_{-0.56}$&$1.22{}^{+0.28}_{-0.55}$&$7.32{}^{+1.48}_{-3.16}$
				\end{tabular}
		\caption{Decay widths (in $\mathrm{GeV}  $) for the semileptonic $\Omega^{+}_{ccb}\to\Xi_{cc}^{++} \ell {\overline{\nu}}_\ell$ transition at different channels.}\label{DECAY}
	\end{ruledtabular}
\end{table}
Assessing  the ratio of decay width in  $\tau$ and $e$/$\mu$ channels is advantageous due to its reduced uncertainty. We find:
\begin{eqnarray}
	R=\frac{\Gamma~[\Omega^{+}_{ccb}\to\Xi_{cc}^{++} \tau~ {\overline{\nu}}_{\tau}]}{\Gamma~[\Omega^{+}_{ccb}\to\Xi_{cc}^{++} e(\mu)~ {\overline{\nu}}_{e(\mu)}]}=0.60^{+0.07}_{-0.02}.
\end{eqnarray}
These results, combined with the mass and residue calculated in our prior work \cite{Najjar:2024deh}, are valuable for refining or finalizing the properties of $\Omega_{ccb}^+$ through future experiments. Comparing our results as  Standard Model (SM) theory  prediction
with future experimental data, could confirm the SM  or open  new  avenue for discovering possible new physics in heavy baryon decays.

\section {conclusion }\label{Sec:five}
So far, there have been no experimental observations of the triply heavy baryons as the last generation of the standard heavy baryons predicted by the quark mode. Few theoretical analysis concerning the weak decay of this group of baryons have been conducted, often focusing on the decay processes involving triply heavy baryons with spin 3/2 transitioning to other triply heavy baryons, or those with spin 1/2 decaying into triply heavy baryons with spin 3/2. This study represents a pioneering investigation into the decay of triply heavy particles with spin 1/2 serving as both initial and final ones in the all lepton channels. In this work, we employed the standard   QCD sum rule method to compute the relevant correlation function, which here is the three-point correlation function. This computation was  conducted first on the physical side in terms of the  vector and axial vector form factors. Subsequently, we derived the three-point correlation function on the QCD side through a series of manual algebraic relations in terms of fundamental QCD parameters like the quark masses, condensation of quarks, gluons, and their combinations, the strong
coupling constant and $\cdots$. In the final step, the form factors were calculated by assuming quark-hadron duality and applying  Borel transformations and matching the same Lorentz structures from both sides. Upon completion of the form factor calculations, we were positioned to determine the decay width of the semileptonic transformation  $ \Omega^{+}_{ccb}\rightarrow \Xi^{++}_{cc}  ~{\ell}\bar\nu_{\ell}$ across all lepton channels. The order of exclusive widths indicate that these channels are accessible in near future. 

The results presented in this study and those on the spectroscopic properties of the triply heavy baryons in Ref.  \cite{Najjar:2024deh} can help experimental groups at different colliders in their search for these interesting heavy particles.  New developments in the experimental side tempt us to believe that we may not be far from observing such states at various hadron colliders.
	\section*{ACKNOWLEDGMENTS}
K. Azizi is grateful to the CERN-TH division for their support and warm hospitality.

\section*{APPENDIX A: DOUBLE SPECTRAL DENSITY}
In this appendix,  we present an alternative procedure for a simple term in QCD side,  as an example, to derive a two-variable dependent spectral density.   For this aim,   we start with
\begin{eqnarray}\label{exampleterm}
	\Pi=\int d^4k_1\int d^4k_2\int d^4k_3 \int d^4x e^{i(k_1+k_2+k_3-p).x}\int d^4y e^{i(-k_1-k_2+p').y} \frac{1} {(k_1^2-m^2_c)^2(k_2^2-m^2_{c})(k_3^2-m^2_b)y^{2}}.
\end{eqnarray}
Using the identity presented in Eq. (\ref{intyx}), we get 
 \begin{eqnarray}\label{exampleterm3}
&\Pi= C \int d^4t&	\int d^4k_1\int d^4k_2\int d^4k_3 \int d^4x e^{i(k_1+k_2+k_3-p).x}\int d^4y e^{i(-k_1-k_2+p'-t).y}\notag\\ &\times&\frac{1} {(k_1^2-m^2_c)^2(k_2^2-m^2_{c})(k_3^2-m^2_b)}\Big(-\frac{1}{t^2}\Big),
\end{eqnarray}
where $ C=\frac{i}{4\pi^2} $ is a constant.  Now, we use 
\begin{eqnarray}\label{fourier}
	\int d^4x e^{i(k_1+k_2+k_3-p).x}\int d^4ye^{i(-k_1-k_2+p'-t).y}= (2\pi)^4\delta^4(k_1+k_2+k_3-p) (2\pi)^4\delta^4(-k_1-k_2+p'-t),
\end{eqnarray}
where the resultant Dirac deltas are used to perform the four-integrals over $k_3$ and $k_2$.
We employ the Feynman parametrization as the body text,  which leads to
\begin{eqnarray}\label{exampleterm4}
&\Pi&=C_1\int d^4t	\int d^4k_1 \int_0^1\int _0^{1-r}\int_0^{1-r-z}dr  dz dv \notag\\ &\times&\frac{r} {[r(k_1^2 - 
    m_c^2) +z(-m_c^2 + (-k_1 + p' - t)^2) +v (-m_b^2 + (p - p' + 
      t)^2) +(1 - r - z - v)t^2]^5},
\end{eqnarray}

where $C_1=i (2\pi)^6 $.  In the next step,  we use the  identity presented in Eq. (\ref{Int}) to perform the remaining integrals, 
which leads to
\begin{eqnarray}\label{exampleterm5}
&\Pi&=C_1 \int_0^1\int_ 0^{1-r}\int_ 0^{1-r-z}dr  dz dv~ \frac{r} {\Delta},
\end{eqnarray}
where
\begin{eqnarray}\label{exampleterm6}
\Delta&=&(-1 + r) \Big[p'^2 r z A_1-\Big(m_c^2 (r +z) A_2+v\big[p^2 r z A_2^2+m_b^2 A_2-q^2 (r + z )A_1\big]\Big)\Big]
\end{eqnarray}		
and
\begin{eqnarray}\label{exampleterm7}
	A_1 &=&\frac{(-1 + r + v +z)}{(r^2 + r (-1 + z) + (-1 + z) z)^2},\nonumber\\
			A_2 &=&\frac{1}{(r^2 + r (-1 +z) + (-1 + z) z)}.						
\end{eqnarray}
Applying double Borel transformation with respect to the $p^2$ 
 and $p'^2$ using the
 \begin{eqnarray}\label{exampleterm8}
 &{\cal B}(M'^2)\frac{1}{\alpha'-p'^2}&=e{^\frac{-\alpha' }{M'^2}},\nonumber\\
&  {\cal B}(M^2)e^{-\beta p^2}&=\delta(1/M^2-\beta),
\end{eqnarray} 
we get
  \begin{eqnarray}\label{exampleterm9}
&{\cal B}(M^2){\cal B}(M'^2)\Pi&=C_1  \int_0^1\int _0^{1-r}\int_ 0^{1-r-z}dr  dz dv ~ B_1~ e^{\frac{\alpha} {M'^2}} \delta(\frac{1}{M^2}-\frac{\beta}{M'^2}),
\end{eqnarray}
where $\alpha=m^2_c B_2+m_b^2B_3-q^2B_4$ and  $\beta=\frac{v }{(-1+r+v+z)}$.  The variables $ B_1$,  $ B_2$ ,  $ B_3$ and $ B_4$ are defined by
\begin{eqnarray}\label{exampleterm10}
	B_1 &=&\frac{1}{(-1 + r)  zA_1},\nonumber\\							B_2 &=&\frac{ (r + z) (r^2 + r (-1 +z) + (-1 + z) z)}{r z (-1 +r + v + z)},\nonumber\\
	B_3&=&\frac{(r^2 + r(-1 +z) + (-1 + z) z) v}{r z (-1 + r + v + z)},\nonumber\\
			B_4&=&\frac{(r + z) v (-1 +r + z + v)}{r z (-1 + r + v + z)}.		
\end{eqnarray}
Performing  integral over $v$,  one obtains
\begin{eqnarray}\label{exampleterm11}
&{\cal B}(M^2){\cal B}(M'^2)\Pi&=C_1  \int_0^1\int _0^{1-r}dr  ~dz~ D_4~\frac{M^2}{M^2+M'^2}~e^{\frac{q^2 D_1} {M'^2+M^2}} e^{\frac{-m_b^2 D_2} {M^2}}e^{\frac{-m_c^2 D_3} {M_1^2}} ,
\end{eqnarray}
where $M_1^2=\frac{M'^2 M^2 }{M'^2+M^2}$ and we  have applied the following shorthand notations:
\begin{eqnarray}\label{exampleterm12}
			D_1 &=&\frac{  (-1 + r + z) (r + z)}{r z },\nonumber\\
	D_2 &=&\frac{-(r^2 +r (-1 + z) + (-1 + z) z)}{r z },\nonumber\\
			D_3 &=&\frac{(r + z) (r^2 +r (-1 +z) + (-1 + z) z)}{r z (-1 + r + z)}	,\nonumber\\
						D_4&=&  \frac{(-1 + r + z )}{(r^2 + r (-1 +z) + (-1 + z) z)^2 (-1 + r)  z} 	.		
\end{eqnarray}
In the last step, we discuss how contribution of the continuum and higher states are subtracted  (see also \cite{Beilin:1984pf,Aliev:2011ufa,Azizi:2018duk}).
To this end, we consider a generic term of the form:
\begin{eqnarray}\label{exampleterm13}
A= \int_0^1\int _0^{1-r}dr ~dz~\frac{M^2}{M^2+M'^2}~e^{\frac{-m_b^2 D_2} {M^2}}e^{\frac{-m_c^2 D_3} {M_1^2}} .
\end{eqnarray}
at $ q^2=0 $.  We introduce new variables by defining  $\sigma_1=\frac{1}{M^2}$ and $\sigma_2=\frac{1}{M'^2}$.  We have
 \begin{eqnarray}\label{exampleterm15}
A&= &(\frac{\sigma_2} {\sigma_1+\sigma_2})~\int_0^1\int _0^{1-r}dr  ~dz~e^{-(m_b^2 D_2+m_c^2 D_3) \sigma_1 }e^{-m_c^2 D_3  \sigma_2} \nonumber\\
&=&\sigma_2~\int_0^1\int _0^{1-r}dr  dz~e^{-(m_b^2 D_2+m_c^2 D_3) \sigma_1 }e^{-m_c^2 D_3  \sigma_2} \int_0^\infty d\xi e^{-\xi(\sigma_1+ \sigma_2)} \nonumber\\
&=&\sigma_2~\int_0^1\int _0^{1-r}dr  dz~\int_0^\infty d\xi   e^{-(m_b^2 D_2+m_c^2 D_3+ \xi) \sigma_1 }e^{-(m_c^2 D_3 + \xi) \sigma_2} \nonumber\\
&=&~-\int_0^1\int _0^{1-r} dr  dz~\int_0^\infty d\xi  \Bigg((\frac{d}{d\xi}) e^{-(m_c^2 D_3+ \xi)\sigma_2}\Bigg)~e^{-(m_b^2 D_2+m_c^2 D_3+ \xi) \sigma_1 }.
\end{eqnarray}
We obtain the double  spectral density by performing another double Borel transformation with respect to $\sigma_1\rightarrow\frac{1}{s}$ and $\sigma_2\rightarrow\frac{1}{s'}$ 
 \begin{eqnarray}\label{exampleterm16}
\rho(s,s')&= &-\int_0^1\int _0^{1-r} dr dz \int_0^\infty d\xi  \Bigg(\frac{d}{d\xi}\delta[s'-(\xi+m_c^2 D_3)]\Bigg)\nonumber\\
&\times&\delta[s-(\xi+ m_b^2 D_2+m_c^2 D_3) ].
\end{eqnarray} 
Performing the $\xi$ integral, finally  the following expression for the double spectral density is found
\begin{eqnarray}\label{exampleterm17}
\rho(s,s')&=&-\int_0^1\int _0^{1-r} dr  dz~ \Bigg(\frac{d}{ds}\delta(s'-s+ m_b^2 D_2)\Bigg) ~\theta(s-( m_b^2 D_2+m_c^2 D_3)),
\end{eqnarray} 
where $ \theta $ is the  Heaviside Theta function. Now, we use this spectral density to write the subtracted correlation function corresponding to the considered term in Borel scheme as:
\begin{eqnarray}\label{exampleterm18}
\Pi^{sub}=\int_{s_L}^{s_0}ds\int_{s'_L}^{s'_0}ds'~\rho(s,s')~e^{-s/M^2}e^{-s'/M'^2}.
\end{eqnarray}
where $s_L = (m_b + 2m_c)^2 $ and $s'_L = ( 2m_c)^2 $.  From the same manner, we can generalized the calculations to the case $ q^2\neq0 $ and write: 
\begin{eqnarray}\label{exampleterm19}
\Pi^{'sub}=\int_{s_L}^{s_0}ds\int_{s'_L}^{s'_0}ds'~\rho(s,s',q^2)~e^{-s/M^2}e^{-s'/M'^2},
\end{eqnarray}
which is the double Borel performed form of the  two-variable spectral representation of the related correlation function.
\section*{APPENDIX B: OPE RESULTS FOR ONE OF THE INVOLVED  STRUCTURES}

In this appendix, we depict the explicit forms of spectral densities $\rho^i(s,s',q^2)$ and functions $\Gamma^i(p^2,p'^2,q^2)$ corresponding to  the structure $\gamma_{\mu} \gamma_{5}$:
\begin{equation}
	\begin{split}
		\rho^{pert}_{\gamma_{\mu} \gamma_5}(s,s',q^2)&=\frac{1}{\sqrt{2}\,512 \pi ^4}
		\int^{1}_{0} du \int^{1-u}_{0} dv \int^{1-u-v}_{0} dz\ \frac{ 3\,D(s,s',q^2)\,\Theta\Big[D(s,s',q^2)\Big]}{K_2\,W_1\,W_2 W_3^5}\\
		&\times \Bigg\{\bigg[D(s,s',q^2) W_2 \Big(2 u W_3 (W_3 + 15 v) - 
		v (W_3 (2 + 11 K_2 W_3)
		+ 9 W_3 v + 11 v^2)\Big)\\
		& + 
		2 K_1 W_3 \Big((W_3 + v) (-3 q^2 W_1 + s' (8 u W_3 + 23 u v + 8 W_4 v) \\
		&+ 
		3 s (u W_3 + v (W_3 + v))) - (3 q^2 - 3 s - 
		8 s') (W_3 u + (4 u + W_4) v) z\Big)
		\bigg]\\
		&-8\beta   \bigg[3 D(s,s',q^2) W_2 \Big(u W_3 (W_3 - 2 v) + 
		v (W_3 (-1 + 3 K_2 W_3) + 4 W_3 v + 3 v^2)\Big)\\
		& + 
		4 K_1 \Big((W_3 + v) \big[s (W_1 - u W_1) + q^2 W_1 W_3 + 
		s' v (u - u^2 + 2 K_2 W_3^2 + 2 (u + W_4) v)\big]\\
		& + \big[(q^2 - 
		s) u W_3^2 + 
		W_3 (-K_2 s' + (-2 + K_2) s' u + s (1 + K_2 - (2 + K_2) u)\\
		& + 
		q^2 (-1 + 2 u + K_2 W_3)) v + (2 q^2 - 2 s + 
		s') W_3 v^2 + (q^2 - s + s') v^3\big] z\Big)\bigg]\\
		&	+	\beta^2\bigg[D(s,s',q^2) W_2 \Big(2 u W_3 (W_3 + 15 v) - 
		v (W_3 (2 + 11 K_2 W_3) + 9 W_3 v + 11 v^2)\Big) \\
		&+ 
		2 K_1 W_3 \Big((W_3 + v) \Big[-3 q^2 (u (u + W_4) + W_4 v) + 
		s' (8 u W_3 + 23 u v + 8 W_4 v) + 
		3 s (u W_3 + v (W_3 + v))\Big] \\
		&- (3 q^2 - 3 s - 
		8 s') \Big[W_3 u + (4 u + W_4) v\Big] z\Big)\bigg]
		\Bigg\},
	\end{split}
\end{equation}
\begin{eqnarray} \label{Rho3}
	&&\rho^3_ {\gamma_{\mu} \gamma_5}(s,s',q^2)=
	\int_{0}^{1}du \int_{0}^{1-u}dv ~\frac{ (-1+\beta)^2 m_b  \langle \bar{u} u\rangle }{8 \sqrt{2}   \pi^2  (W_5)^3}\, \Theta\Big[L(s,s',q^2)\Big],
\end{eqnarray}
\begin{eqnarray}\label{Rho4}
	&&\rho^4_{\gamma_{\mu} \gamma_5}(s,s',q^2)=\int_{0}^{1}du \int_{0}^{1-u}dv \int_{0}^{1-u-v}dz\frac{\Theta\Big[D(s,s',q^2)\Big]}{12288\, \sqrt{2} K_2^3 W_1  W_3^7 \pi^2}\Big\langle\frac{\alpha_{s}GG}{\pi}\Big\rangle\  \notag\\
	&& \times \Bigg\{\bigg[3 \Big(-85 u^2 W_3^3 + u W_3^2 (129 - 333 u + 44 K_2 W_3) v - 
	W_3 (44 + 8 K_2 W_3 + u (-486 + 585 u + 16 K_2 W_3)) v^2 \notag\\
	&& + 
	W_3 (80 - 373 u + 8 K_2 W_3) v^3 + 4 (7 - 13 u) v^4 + 
	8 v^5\Big) +\Big(24 u (-25 + 21 u) W_3^2 + 
	3 W_3 (200 - 110 K_2 W_3  \notag\\
	&&+ u (-648 + 580 u + 95 K_2 W_3)) v + 
	W_3 (-1158 + 765 u + 242 K_2 W_3) v^2 + 5 (-160 + 151 u) v^3 + 
	242 v^4\Big) z \notag\\
	&& + 
	6 \Big(2 u W_3 (50 W_3 + 101 v) + 
	5 v (W_3 (-20 + 11 K_2 W_3) + 31 W_3 v + 11 v^2)\Big) z^2\bigg]\notag\\
	&&+\beta\bigg[96 \Big(-u^5 + u^4 (3 + (-9 + K_2) v) + 
	u^3 (-3 + (21 + K_2 (-3 + v) - 22 v) v) + 
	2 W_4 v^2 (-1 + K_2 + v^2)\notag\\
	&& + 
	u W_4 v (-3 + K_2 - 4 K_2 v + v (18 + v)) + 
	u^2 \Big[1 + 
	v \big(-15 + (41 - 15 v) v + K_2 (3 + 2 (-2 + v) v)\big)\Big]\Big)\notag\\
	&& +\big(-W_3^2 u (11 + 18 u) + (3 K_2 W_3^2 (-20 + 13 u) + 
	u (-30 + (84 - 43 u) u) + 
	11 W_4) v\notag\\
	&& + (58 K_2 W_3^2 + (27 - 85 u) u - 47 W_4) v^2 + 
	2 (42 u + 29 W_4) v^3\big) z\notag\\
	&& + \big(11 W_3^2 u + (60 K_2 W_3^2 - 
	u (1 + 10 u) - 11 W_4) v + (71 u + 60 W_4) v^2\big) z^2\bigg]\notag\\
	&&+\beta^2\bigg[-282 W_3^3 u^2 + 
	6 (57 + 17 K_2 W_3 - 127 u) W_3^2 u v + 
	2 \big[u \big(-468 + 933 u - 495 u^2 + K_2 (331 + u (-410 + 163 u))\big)\notag\\
	&& + 
	84 K_2 u4 + 30 u4^2\big] v^2+ 
	8 \big((20 + 21 K_2 (-2 + u) - 26 u) u + 21 W_4^2\big) v^3 + 
	434 u v^4 \notag\\
	&& + \big[9 W_3^2 u (-18 + 
	5 u) + (-K_2 W_3^2 (-38 + 79 u) + 
	u (499 + u (-755 + 418 u)) + 162 W_4) v \notag\\
	&&+ 
	2 (27 K_2 W_3^2 + (319 - 303 u) u + 97 W_4) v^2 + (-181 u + 
	54 W_4) v^3\big] z \notag\\
	&&+ 
	2 \big(81 W_3^2 u - (19 K_2 W_3^2 + (305 - 224 u) u + 
	81 W_4) v + (62 u - 19 W_4) v^2\big) z^2\bigg]\Bigg\} ,
\end{eqnarray}
and
\begin{eqnarray}\label{Rho5}
	&&\Gamma^5_{\gamma_{\mu} \gamma_5}(p^2,p'^2,q^2)=
	\frac{\langle \bar{u}u\rangle m_o^2 }{32\, \sqrt{2}\, \pi^2 W_5^7\,L^3_1(p^2,p'^2,q^2)}\notag\\
	&&\times\Bigg\{(-1+\beta)^2\Big[\Big(- m_b  u  v (L_1(p^2,p'^2,q^2) (L_1(p^2,p'^2,q^2) - 2 m_c^2) W_5^2 + 
	2 m_c^2 u v p'^2)\Big)\Big]\notag\\
	&&
	-\Big[(1 + \beta)  u  v  p'^2 \Big(2 (-1 + \beta) L_1(p^2,p'^2,q^2) m_c W_5^4 + 
	u W_5 \big((-1 + \beta) m_c (p^2 + p'^2 - q^2) u\notag\\
	&& - 
	2 (1 + \beta) L_1(p^2,p'^2,q^2) m_b W_5\big) v + (1 + \beta) m_b u^2 v^2 p'^2\Big)\,\frac{2}{W_5^2}\Big]\notag\\
	&&+\Big[(-1 + \beta) (1 + 
	\beta) m_c  \big(L_1(p^2,p'^2,q^2) W_5 (3 (p^2 + p'^2 - q^2) u^2 v + 
	L_1(p^2,p'^2,q^2) W_5^2 (u + 2 v)) + 
	2 u^2 v^2 p'^4\big)\Big]\Bigg\},
\end{eqnarray}
where we have defined:
\begin{eqnarray}
	D(s,s',q^2) &=&\Big(\frac{W_3 (m_c^2 W_1 W_5 - s' u W_2 v + (m_b^2 W_1 - q^2 W_2 W_5 + s u v) z)}{W_1^2}  \Big),\nonumber\\
	L(s,s',q^2) &=&\Big(-m_c^2 +\frac{(m_b^2 - q^2)  W_5 W_6 + s' u v}{W_5^2}  \Big),\nonumber\\
	L_1(p^2,p'^2,q^2) &=&\Big(-m_c^2 +\frac{(m_b^2 - q^2)  W_5 W_6 + p'^2 u v}{W_5^2}  \Big),\nonumber\\
	K_1&=&\frac{uv W_2 W_3}{W_1^2},\nonumber\\
	K_2&=&\frac{-W_1}{W_3^2}.
\end{eqnarray}
Here, we  have applied the following shorthand symbolizations, as well: 
\begin{eqnarray}
	W_1 &=&(u^2 + u (-1 + v) + (-1 + v) v),\nonumber\\
	W_2 &=& (-1 + u + v + z),\nonumber\\
	W_3&=& (-1+u),\nonumber\\
	W_4&=&(-1+v),\nonumber\\
	W_5&=&(u+v),\nonumber\\
	W_6&=&(-1+u+v).
\end{eqnarray}

	
	\end{document}